\documentclass[sigconf]{acmart}

\AtBeginDocument{%
  \providecommand\BibTeX{{%
    \normalfont B\kern-0.5em{\scshape i\kern-0.25em b}\kern-0.8em\TeX}}}

\copyrightyear{2024}
\acmYear{2024}
\setcopyright{acmlicensed}\acmConference[CHI '24]{Proceedings of the CHI Conference on Human Factors in Computing Systems}{May 11--16, 2024}{Honolulu, HI, USA}
\acmBooktitle{Proceedings of the CHI Conference on Human Factors in Computing Systems (CHI '24), May 11--16, 2024, Honolulu, HI, USA}
\acmDOI{10.1145/3613904.3642830}
\acmISBN{979-8-4007-0330-0/24/05}

\usepackage[most]{tcolorbox} % Custom badges
\usepackage{xcolor} % Custom color definitions
\usepackage{listings} % For code blocks
\usepackage{soul} % for strikeouts in revision

\newcommand\mystrut{\rule[-1pt]{0pt}{1pt}} % Standardizes badge height
\newcommand{\lloomURL}{\url{https://github.com/michelle123lam/lloom}}

\newtcbox{\badge}[1][]{
  on line, 
  arc=2pt,
  colback=#1,
  colframe=#1,
  fontupper=\color{black}\ttfamily\mystrut,
  boxrule=1pt, 
  boxsep=0pt,
  left=2pt,
  right=2pt,
  top=1pt,
  bottom=1pt,
}  % Defines badge style

\definecolor{bl}{HTML}{C0E0FD} % blue
\definecolor{rd}{HTML}{F4C7CF} % red
\definecolor{yl}{HTML}{FDE5C0} % yellow
\definecolor{gr}{HTML}{D9D9D9} % grey

\definecolor{maroon}{cmyk}{0, 0.87, 0.68, 0.32}
\definecolor{halfgray}{gray}{0.55}
\definecolor{ipython_frame}{RGB}{207, 207, 207}
\definecolor{ipython_bg}{RGB}{247, 247, 247}
\definecolor{ipython_red}{RGB}{186, 33, 33}
\definecolor{ipython_green}{RGB}{0, 128, 0}
\definecolor{ipython_cyan}{RGB}{64, 128, 128}
\definecolor{ipython_purple}{RGB}{170, 34, 255}
\lstdefinelanguage{Markdown}{
    basicstyle=\ttfamily\footnotesize,
}
\lstdefinelanguage{iPython}{
    morekeywords={access,and,break,class,continue,def,del,elif,else,except,exec,finally,for,from,global,if,import,in,is,lambda,not,or,pass,print,raise,return,try,while},%
    morekeywords=[2]{abs,all,any,basestring,bin,bool,bytearray,callable,chr,classmethod,cmp,compile,complex,delattr,dict,dir,divmod,enumerate,eval,execfile,file,filter,float,format,frozenset,getattr,globals,hasattr,hash,help,hex,id,input,int,isinstance,issubclass,iter,len,list,locals,long,map,max,memoryview,min,next,object,oct,open,ord,pow,property,range,raw_input,reduce,reload,repr,reversed,round,set,setattr,slice,sorted,staticmethod,str,sum,super,tuple,type,unichr,unicode,vars,xrange,zip,apply,buffer,coerce,intern},%
    sensitive=true,%
    morecomment=[l]\#,%
    morestring=[b]',%
    morestring=[b]",%
    morestring=[s]{'''}{'''},% used for documentation text (mulitiline strings)
    morestring=[s]{"""}{"""},% added by Philipp Matthias Hahn
    morestring=[s]{r'}{'},% `raw' strings
    morestring=[s]{r"}{"},%
    morestring=[s]{r'''}{'''},%
    morestring=[s]{r"""}{"""},%
    morestring=[s]{u'}{'},% unicode strings
    morestring=[s]{u"}{"},%
    morestring=[s]{u'''}{'''},%
    morestring=[s]{u"""}{"""},%
    literate=
    {á}{{\'a}}1 {é}{{\'e}}1 {í}{{\'i}}1 {ó}{{\'o}}1 {ú}{{\'u}}1
    {Á}{{\'A}}1 {É}{{\'E}}1 {Í}{{\'I}}1 {Ó}{{\'O}}1 {Ú}{{\'U}}1
    {à}{{\`a}}1 {è}{{\`e}}1 {ì}{{\`i}}1 {ò}{{\`o}}1 {ù}{{\`u}}1
    {À}{{\`A}}1 {È}{{\'E}}1 {Ì}{{\`I}}1 {Ò}{{\`O}}1 {Ù}{{\`U}}1
    {ä}{{\"a}}1 {ë}{{\"e}}1 {ï}{{\"i}}1 {ö}{{\"o}}1 {ü}{{\"u}}1
    {Ä}{{\"A}}1 {Ë}{{\"E}}1 {Ï}{{\"I}}1 {Ö}{{\"O}}1 {Ü}{{\"U}}1
    {â}{{\^a}}1 {ê}{{\^e}}1 {î}{{\^i}}1 {ô}{{\^o}}1 {û}{{\^u}}1
    {Â}{{\^A}}1 {Ê}{{\^E}}1 {Î}{{\^I}}1 {Ô}{{\^O}}1 {Û}{{\^U}}1
    {œ}{{\oe}}1 {Œ}{{\OE}}1 {æ}{{\ae}}1 {Æ}{{\AE}}1 {ß}{{\ss}}1
    {ç}{{\c c}}1 {Ç}{{\c C}}1 {ø}{{\o}}1 {å}{{\r a}}1 {Å}{{\r A}}1
    {€}{{\EUR}}1 {£}{{\pounds}}1
    {^}{{{\color{ipython_purple}\^{}}}}1
    {=}{{{\color{ipython_purple}=}}}1
    {+}{{{\color{ipython_purple}+}}}1
    {*}{{{\color{ipython_purple}$^\ast$}}}1
    {/}{{{\color{ipython_purple}/}}}1
    {+=}{{{+=}}}1
    {-=}{{{-=}}}1
    {*=}{{{$^\ast$=}}}1
    {/=}{{{/=}}}1,
    literate=
    *{-}{{{\color{ipython_purple}-}}}1
     {?}{{{\color{ipython_purple}?}}}1,
    identifierstyle=\color{black}\ttfamily,
    commentstyle=\color{ipython_cyan}\ttfamily,
    stringstyle=\color{ipython_red}\ttfamily,
    keepspaces=true,
    showspaces=false,
    showstringspaces=false,
    basicstyle=\ttfamily\footnotesize,
    keywordstyle=\color{ipython_green}\ttfamily,
}

\begin{document}

%%
%% The "title" command has an optional parameter,
%% allowing the author to define a "short title" to be used in page headers.
\title{Concept Induction: Analyzing Unstructured Text with High-Level Concepts Using LLooM}

%%
%% The "author" command and its associated commands are used to define
%% the authors and their affiliations.
%% Of note is the shared affiliation of the first two authors, and the
%% "authornote" and "authornotemark" commands
%% used to denote shared contribution to the research.
\author{Michelle S. Lam}
\orcid{0000-0002-3448-5961}
\affiliation{%
  \institution{Stanford University}
  \city{Stanford}
  \state{CA}
  \country{USA}
}
\email{mlam4@cs.stanford.edu}

\author{Janice Teoh}
\orcid{0009-0002-7550-7300}
\affiliation{%
  \institution{Stanford University}
  \city{Stanford}
  \state{CA}
  \country{USA}
}
\email{jteoh2@stanford.edu}

\author{James A. Landay}
\orcid{0000-0003-1520-8894}
\affiliation{%
  \institution{Stanford University}
  \city{Stanford}
  \state{CA}
  \country{USA}
}
\email{landay@stanford.edu}

\author{Jeffrey Heer}
\orcid{0000-0002-6175-1655}
\affiliation{%
  \institution{University of Washington}
  \city{Seattle}
  \state{WA}
  \country{USA}
}
\email{jheer@uw.edu}

\author{Michael S. Bernstein}
\orcid{0000-0001-8020-9434}
\affiliation{%
  \institution{Stanford University}
  \city{Stanford}
  \state{CA}
  \country{USA}
}
\email{msb@cs.stanford.edu}

%%
%% By default, the full list of authors will be used in the page
%% headers. Often, this list is too long, and will overlap
%% other information printed in the page headers. This command allows
%% the author to define a more concise list
%% of authors' names for this purpose.
\renewcommand{\shortauthors}{M.S. Lam, J. Teoh, J.A. Landay, J. Heer, M.S. Bernstein}

%%
%% The abstract is a short summary of the work to be presented in the
%% article.
\begin{abstract}
  Data analysts have long sought to turn unstructured text data into meaningful concepts. Though common, topic modeling and clustering focus on lower-level keywords and require significant interpretative work. We introduce \textit{concept induction}, a computational process that instead produces high-level concepts, defined by explicit inclusion criteria, from unstructured text. For a dataset of toxic online comments, where a state-of-the-art BERTopic model outputs ``women, power, female,'' concept induction produces high-level concepts such as ``Criticism of traditional gender roles'' and ``Dismissal of women's concerns.'' We present LLooM, a concept induction algorithm that leverages large language models to iteratively synthesize sampled text and propose human-interpretable concepts of increasing generality. We then instantiate LLooM in a mixed-initiative text analysis tool, enabling analysts to shift their attention from interpreting topics to engaging in theory-driven analysis. Through technical evaluations and four analysis scenarios ranging from literature review to content moderation, we find that LLooM’s concepts improve upon the prior art of topic models in terms of quality and data coverage. In expert case studies, LLooM helped researchers to uncover new insights even from familiar datasets, for example by suggesting a previously unnoticed concept of attacks on out-party stances in a political social media dataset.

% Data analysts have long sought to turn unstructured text data into meaningful concepts. Though common, topic modeling and clustering focus on lower-level keywords and require significant interpretative work. We introduce concept induction, a computational process that instead produces high-level concepts, defined by explicit inclusion criteria, from unstructured text. For a dataset of toxic online comments, where a state-of-the-art BERTopic model outputs “women, power, female,” concept induction produces high-level concepts such as “Criticism of traditional gender roles” and “Dismissal of women's concerns.” We present LLooM, a concept induction algorithm that leverages large language models to iteratively synthesize sampled text and propose human-interpretable concepts of increasing generality. We then instantiate LLooM in a mixed-initiative text analysis tool, enabling analysts to shift their attention from interpreting topics to engaging in theory-driven analysis. Through technical evaluations and four analysis scenarios ranging from literature review to content moderation, we find that LLooM’s concepts improve upon the prior art of topic models in terms of quality and data coverage. In expert case studies, LLooM helped researchers to uncover new insights even from familiar datasets, for example by suggesting a previously unnoticed concept of attacks on out-party stances in a political social media dataset.

\end{abstract}

%%
%% The code below is generated by the tool at http://dl.acm.org/ccs.cfm.
%% Please copy and paste the code instead of the example below.
%%


%%
%% Keywords. The author(s) should pick words that accurately describe
%% the work being presented. Separate the keywords with commas.

%% A "teaser" image appears between the author and affiliation
%% information and the body of the document, and typically spans the
%% page.
\begin{teaserfigure}
  \includegraphics[width=\textwidth]{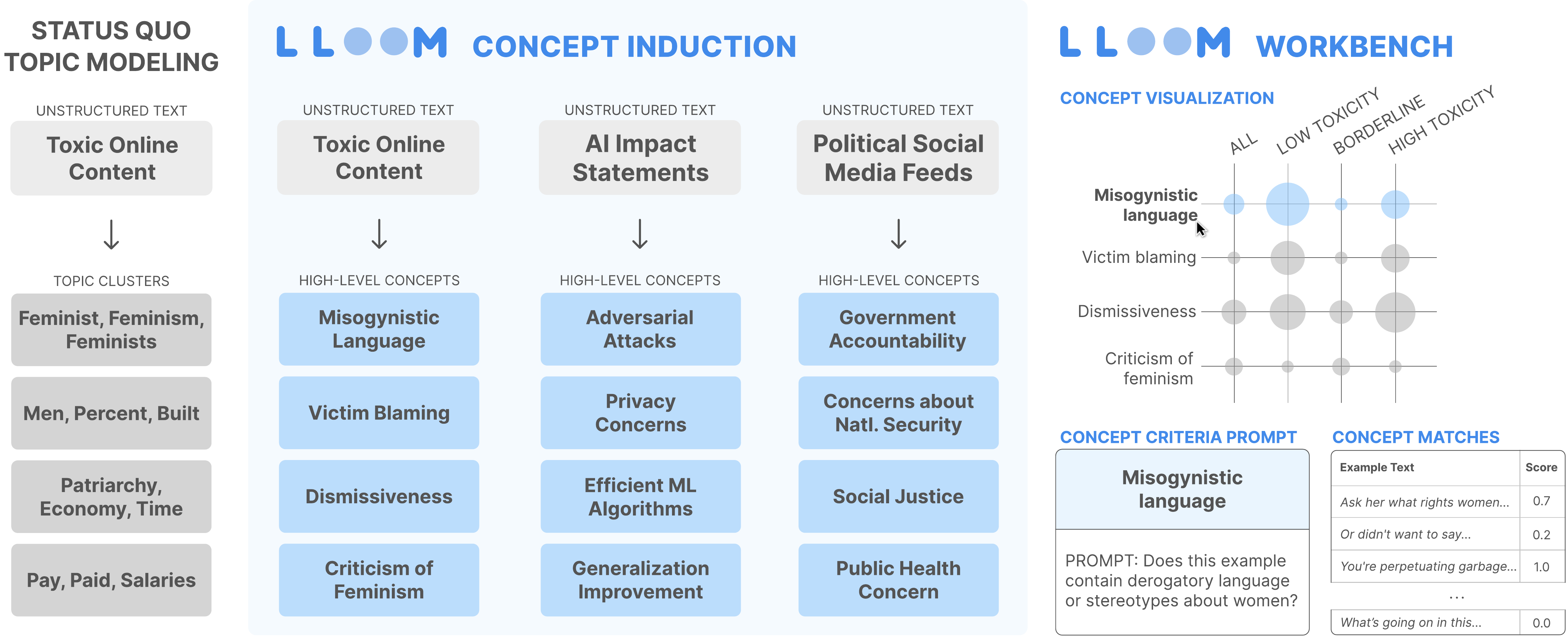}
  \caption{
    \textbf{A summary of the LLooM concept induction algorithm.} Status quo topic models tend to produce topics aligned with low-level keywords (e.g., ``feminist, feminism''). We introduce LLooM, a \textit{concept induction} algorithm that takes in unstructured text and produces high-level concepts (e.g., ``Criticism of Feminism'') defined by explicit \textit{inclusion criteria}. We instantiate this algorithm in the LLooM Workbench, a mixed-initiative text analysis tool that can amplify the work of analysts by automatically visualizing datasets in terms of interpretable, high-level concepts. 
  }
  \Description{The left half of the figure summarizes the LLooM concept induction technique by contrasting status quo topic modeling (with topic clusters like ``feminist, feminism'') and LLooM concept induction, which produces concepts like Misogynistic Language, Victim Blaming, Dismissiveness. The right half summarizes the LLooM Workbench, which has a matrix visualization for concepts, concept criteria prompt, and a table of concept matches based on applying criteria to examples.}
  \label{fig:pull_figure}
\end{teaserfigure}

% \received{20 February 2007}
% \received[revised]{12 March 2009}
% \received[accepted]{5 June 2009}

%%
%% This command processes the author and affiliation and title
%% information and builds the first part of the formatted document.
\maketitle

\section{Introduction}
Much of the world's information is bound up in unstructured text, but it is challenging to make sense of this data. Topic modeling algorithms---such as Latent Dirichlet Allocation (LDA) and unsupervised clustering based on language model embeddings such as BERTopic---have become ubiquitous tools for wading through large-scale, unstructured data~\cite{blei2003latent,reimers2019sentencebert}. Spreading to domains like social science and medicine, topic models have had far-reaching impact: researchers have used these models to analyze scientific abstracts, social media feed content, and historical newspaper coverage in order to investigate phenomena like scientific research trends, political polarization, public health measures, and media framing~\cite{griffiths2004finding, ramage2010twitter, dimaggio2013exploiting,demszky2019analyzing,tsur2015frame,paul2011you}.

However, the topics produced by these models are defined relative to low-level text signals such as keywords, requiring substantial effort from the analyst who must interpret, validate, and reason about those topics.
For example, when applied to a dataset of misogynistic social media posts, a state-of-the-art BERTopic model produces competent but low-level topics such as ``women, power, female'' and ``feminists, feminism, feminist,'' which are on-topic but too generic to help an analyst answer questions such as ``how are women in power described?'' and ``what kinds of arguments are levied against feminists?'' 
This gap arises because topic models rely on measures of term co-occurrence or embedding distances, which are highly correlated with low-level textual similarity and are often unreliable proxies for human judgement~\cite{hellrich-hahn-2016-bad, zhou-etal-2022-problems,li-etal-2020-sentence}.
Moreover, topic models often produce topics that are too general, too specific, or that are generally incoherent (``junk'' topics, e.g.,~``morning, snoring, sir'')~\cite{chuang13topicModelDiagnostics,alsumait2009topicSignificance}. Analysts lack recourse when input texts are categorized into uninformative groups.
The tasks that analysts must perform---generating research questions, formulating hypotheses, and producing insights---are dependent on the creation of \textit{high-level concepts}, which we define as human-interpretable descriptions defined by explicit \textit{inclusion criteria}.

In this paper, we introduce \textit{concept induction}, the task of extracting high-level concepts from unstructured text to amplify theory-driven data analysis.
For example, given the same dataset of potentially misogynistic social media posts that the BERTopic model labeled with ``women, power, female'' and ``feminists, feminism, feminist,'' concept induction seeks to identify concepts such as ``Criticism of traditional gender roles'' and ``Dismissal of women's concerns.''
Each concept is defined by detailed criteria in natural language: e.g., ``Does the example critique or challenge traditional gender roles or expectations?'', or ``Does the example dismiss or invalidate women's fears, concerns, or experiences?''. These defining criteria are supported by a set of representative text examples that best demonstrate the idea of the concept, along with concept scores ranging from 0 to 1 that indicate the extent to which every example in the dataset aligns with that concept (Figure~\ref{fig:pull_figure}). 

\begin{figure*}[!tb]
  \includegraphics[width=0.95\textwidth]{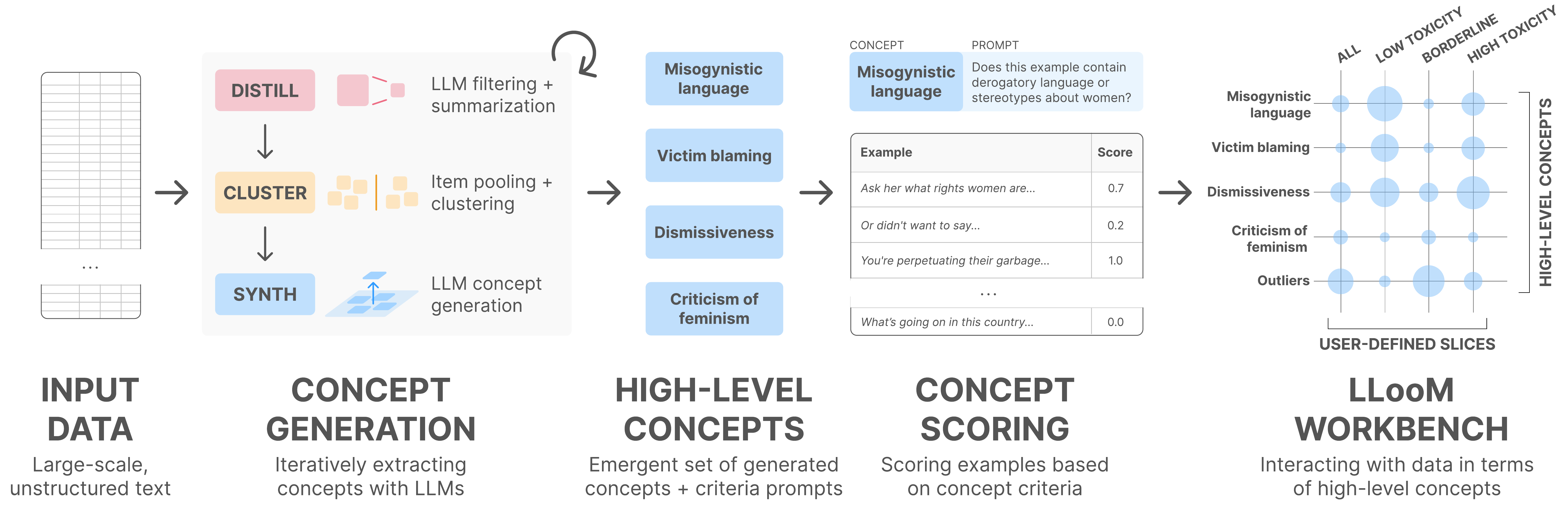}
  \caption{
    \textbf{A process overview of the LLooM concept induction algorithm}. Starting from (1)~unstructured text data, LLooM performs (2)~concept generation aided by an LLM to produce (3)~high-level concepts, which consist of generated natural language descriptions and explicit criteria in the form of zero-shot LLM prompts. LLooM performs (4)~concept scoring based on concept criteria prompts and visualizes data in terms of concepts in the (5)~LLooM Workbench, a mixed-initiative text analysis tool.
  }
  \label{fig:process_overview}
  \Description{The figure proceeds from an image of an input data table leading into a concept generation step that consists of a pipeline of Distill, Cluster, and Synthesize operators that can be looped. This step leads to a set of blocks representing concepts like ``Misogynistic language'' and ``Criticism of feminism''. That leads to a block with one concept block and its associated criteria prompt block, along with a table displaying text examples and corresponding 0-1 concept scores. Finally, that leads to a matrix view with high-level concepts as rows and user-defined slices as columns.}
\end{figure*}

To enable these results, we develop a concept induction algorithm called LLooM, which draws on the ability of large language models (LLMs) like GPT-3.5 and GPT-4~\cite{openai2023gpt4} to generalize from examples: LLooM samples extracted text and iteratively synthesizes proposed concepts of increasing generality (Figure~\ref{fig:process_overview}). 
Once data has been synthesized into a concept, we can move up to the next abstraction level; we can generalize from smaller, lower-level concepts to broader, \textit{higher-level concepts} by repeating the process with concepts as the input. 
Since concepts include explicit inclusion criteria, we can expand the reach of any generated concept to consistently \textit{classify new data} through that same lens and discover gaps in our current concept set.
These core capabilities of synthesis, classification, and abstraction are what allow LLooM to iteratively generate concepts, apply them back to data, and bubble up to higher-level concepts.

Instantiated in a mixed-initiative text analysis tool that we call the LLooM Workbench, our algorithm amplifies the work of analysts by automatically visualizing datasets in terms of interpretable, high-level concepts.
The LLooM Workbench additionally offers analysts a traceable and malleable \textit{process}. Each extracted concept is not just a final label, but can be unrolled into an auditable trace of the lower-level subconcepts that led to the concept (e.g., ``Women's responsibilities,'' ``Traditional gender roles,'' and ``Power dynamics and women'' led to the ``Criticism of traditional gender roles'' concept), where each subconcept is again paired with reviewable criteria and representative examples. Further, analysts can use the LLooM Workbench to seed the algorithm, steering its attention toward particular concepts.

With a series of four analysis scenarios, we first illustrate how LLooM works in practice by comparing it to a state-of-the-art BERTopic model. These scenarios span a variety of domains and analysis goals: 
a content moderation task with a dataset of toxic online content~\cite{kumar2021designing},
an analysis of partisan animosity on social media feeds with a political social media content dataset~\cite{jia2023embedding}, 
a literature review analyzing the industry impact of the field of HCI with paper abstracts from the past 30 years~\cite{cao2023breaking},
and an analysis of anticipated consequences of AI research with a dataset of broader impact statements from NeurIPS 2020~\cite{nanayakkara2021impactStatements}. 
In these scenarios, LLooM not only covers most topics surfaced by BERTopic, but also provides on average $2.0$ times the number of high-quality topics.
Additionally, cluster-based topic models struggle with large sets of uncategorized examples (averaging $77.7\%$ coverage), but LLooM concepts cover on average $93\%$ of examples. 

Then, in a set of technical evaluations, we benchmark LLooM against zero-shot GPT-4 variants and BERTopic for real-world and synthetic datasets; we find that LLooM provides performance gains over baseline methods. These benefits are especially strong for unseen datasets ($p < .02$) and nuanced concepts ($p < .0001$) where baseline methods struggle; LLooM improves ground truth concept coverage by at least 17.9\% and 16.0\% in those cases, respectively.
While both LLooM and GPT-4 can produce overarching, summary-style concepts, LLooM is capable of additionally producing the nuanced and grounded concepts that analysts seek to more richly characterize patterns in data.
In expert case studies, we also gave original researchers for two of the analysis scenarios access to LLooM to re-analyze their data. The researchers used LLooM Workbench to interactively steer concepts and initiate theory-driven explorations (e.g., refining a concept of ``Policy-related'' social media posts to those where policy was \textit{blamed for a crisis}, or drawing on domain knowledge to add a new concept for ``Social distrust'' defined by ``distrust of other people or society'').

LLooM instantiates a novel approach to data analysis that allows analysts to see and explore data in terms of concepts rather than sifting through model parameters. By transforming unstructured data into high-level concepts that analysts can understand and control, LLooM can augment analysts to draw out new insights, weave together connections, and form a narrative tapestry supported by input data.
This paper introduces the following contributions:
\begin{itemize}
    \item \textbf{The LLooM algorithm}. We introduce LLooM, a \textit{concept induction} algorithm that extracts and applies concepts to make sense of unstructured text datasets. LLooM leverages large language models to synthesize sampled text spans, generate concepts defined by explicit criteria, apply concepts back to data, and iteratively generalize to higher-level concepts.
    \item \textbf{The LLooM Workbench}. We instantiate the LLooM algorithm in the LLooM Workbench, a text analysis tool that amplifies \textit{theory-driven data analysis} by allowing users to visualize and interact with text data in terms of high-level concepts. The tool is available in computational notebooks or a standalone Python package.\footnote{Code available at \lloomURL}
    \item \textbf{Evaluation with analysis scenarios, a technical evaluation, and expert case studies}. We present four analysis scenarios and a technical evaluation demonstrating how LLooM enables analysts to derive insights from data that extend beyond status quo tools. LLooM improves upon the quality and coverage of topic models and helps expert analysts to uncover novel insights even on familiar datasets.
\end{itemize}

\section{Related Work}
To instantiate a concept-centered approach for understanding and interacting with data, LLooM draws on prior literature in topic modeling and unsupervised clustering, qualitative analysis, and mixed-initiative data analysis tools.

\subsection{Topic Modeling and Clustering: Automated Concept Development}
\label{section:rw_topic_modeling}
A vast amount of important information exists as large and unstructured text datasets---global social media posts, corpora of historical documents, massive logs of model-generated output---but it is challenging to make sense of this kind of data. Today, many data analysts rely on topic modeling and unsupervised clustering to \textit{automatically} summarize or explore data. 
Latent Dirichlet Allocation (LDA), a classic topic modeling approach, represents documents as distributions over topics and represents topics as distributions over words, and generates latent topics based on the co-occurrence of words in documents~\cite{blei2003latent}. While easy to apply, a persistent issue with LDA is that its topics may be incoherent or irrelevant to the analyst~\cite{chuang13topicModelDiagnostics,alsumait2009topicSignificance,chang2009readingTeaLeaves}. Furthermore, its bag-of-words (or low-dimensional n-gram) assumptions limit topics to simpler ideas that can be captured with keywords.

More recent approaches perform unsupervised clustering on high-dimensional vector embeddings to uncover latent topics without relying directly on keywords. Popular packages like BERTopic~\cite{grootendorst2020bertopic} streamline the common pipeline of embedding text data (e.g., using a pre-trained model like BERT~\cite{devlin2018bert,reimers2019sentencebert}), performing dimensionality reduction, and applying a clustering algorithm (e.g., k-means, agglomerative clustering, HDBSCAN~\cite{mcinnes2017accelerated}) to recover groups of similar examples based on distance metrics. 
Unsupervised clustering loosens the mapping from topics to keywords, but because embedding distances are still highly correlated with low-level text similarity rather than human judgment of semantic similarity, resulting topics frequently align with surface level features~\cite{hellrich-hahn-2016-bad,li-etal-2020-sentence}. While today's topic models appear highly performant based on automated metrics, recent work has highlighted that these metrics may be strongly misaligned with true human evaluations of topic quality~\cite{hoyle2022neuralTopicModels, hoyle2021topicModelEval}---there is still a critical gap between automatically generated topics and meaningful interpretations.
LLooM addresses this gap by supporting a workflow for data analysts to extract interpretable, high-level concepts from unstructured text.

\subsection{Qualitative Analysis: Manual Concept Development}
In contrast to common machine learning approaches, qualitative analysis methods have long acknowledged that data interpretations are varied, subjective, and highly dependent on one's analysis goals~\cite{baumer2017groundedtheory,muller2017ML_groundedTheory}.
Qualitative coding processes, such as grounded theory methods, have the researcher engage in \textit{manually} reviewing and interpreting the data, typically starting from line-by-line, lower-level summaries and proceeding to rounds of thematic grouping and synthesis into codes~\cite{muller2014curiosity, charmaz2006constructing}. 
Once codes have been synthesized, they are applied back to the data in a process of ``constant comparison,'' which both elucidates the data and tests the robustness and richness of the current codes. These synthesized codes also serve as the input for each successive round of coding to derive broader, more abstractive insights.
The LLooM algorithm draws inspiration from qualitative coding processes, seeking to bring the benefits of iterative interpretation, code development, and refinement to automated data analysis tools.

Given the substantial labor involved in conducting qualitative analysis, researchers have explored algorithmic systems that use AI to aid qualitative analysts with both inductive coding (generating codes from data) and deductive coding (applying codes back to data)~\cite{rietz2021cody, drouhard2017aeonium,chen2018ML_QualCoding}. Most recently, research at the intersection of LLMs and qualitative analysis has focused on amplifying deductive coding processes and found that LLMs perform fairly well in coding data with existing codebooks, though not enough for full reliance~\cite{ziems2023large, xiao2023qualitativeLLMs}.
Meanwhile, novel systems designed to aid inductive coding, such as PaTAT~\cite{patat2023} and Scholastic~\cite{hong2022scholastic}, have explored opportunities for human-AI collaboration that keep the inductive code generation work in the hands of human analysts and leverage AI to sample and re-organize data or to formalize themes into decision rules. 
We build on this work to augment analysts who seek to extract meaningful high-level concepts from their data. However, LLooM investigates whether options for \textit{AI-initiated} concept generation can further extend the work of analysts as a tool for thought to reflect on a wider range of potential data analysis directions.

\subsection{AI-Assisted Data Analysis: Mixed-Initiative Concept Development}
Our work builds on a substantial body of mixed-initiative approaches to aid data analysis, and we especially draw attention to prior work that similarly seeks to extract human-interpretable concepts from data.
Work in topic modeling investigated the challenges---such as technical barriers, interpretability, and trust---that social scientists and data analysts encounter when using topic models~\cite{chuang2012interpretationTrust, ramage2009topic,baumer2017groundedtheory}. In the face of uninterpretable topics, researchers found that interactive visual analysis systems such as Termite, LDAvis, and Semantic Concept Spaces could enable analysts to identify coherent themes and build trust in topic models~\cite{chuang2012termite, sievert-shirley-2014-ldavis,Chuang2014ComputerAssistedCA, el2019semantic}. 
LLooM analogously enables analysts to visualize and iterate on model outputs to facilitate interpretability and trust.

Beyond topic modeling, work at the intersection of HCI and AI has assisted data sensemaking by aligning technical abstractions to user-understandable \textit{concepts}. 
Interactive machine learning tools such as FeatureInsight~\cite{brooks2015featureinsight} and AnchorViz~\cite{chen2018anchorviz} help users to build dictionary- or example-based concepts to explore data and improve classifier performance.
Model Sketching leverages LLMs to allow ML practitioners to create sketch-like models by composing human-understandable concepts~\cite{lam2023modelSketching}.
Systems like GANzilla~\cite{evirgen2022ganzilla} and Sensecape~\cite{suh2023sensecape} support sensemaking with generative models by organizing outputs into conceptual groupings that are meaningful to the user, such as system-provided image-editing directions or user-curated hierarchical canvases.
In statistical data analysis, systems like Tisane~\cite{jun2022tisane} aid an often-overlooked process of hypothesis formalization~\cite{jun2022hypothesisFormalization} by allowing analysts to iterate back and forth between conceptual hypotheses and model implementations.

Meanwhile, recent work in NLP has explored how LLMs might aid text analysis by proposing natural language explanations for clusters~\cite{wang2023goaldriven}, augmenting expert demonstrations for semi-supervised text clustering~\cite{viswanathan2023LLM_clustering}, or generating and assigning interpretable topics~\cite{pham2023topicgpt}. 
LLooM builds on the goal of orienting data analysis around human-understandable concepts, but takes a stronger stance about the \textit{requirements}, \textit{scope}, and \textit{application} of extracted concepts. To be most useful for the data analysis tasks of forming hypotheses and answering research questions, we require concepts to be defined by a human-understandable description and explicit inclusion criteria. To support a rich understanding of text, the LLooM algorithm produces concepts at the scope of not just broad topic-level patterns, but also nuanced and specific text attributes. Finally, while the tasks of text clustering and topic modeling focus on producing \textit{outputs} to aid data interpretation, the LLooM Workbench instantiates concepts as \textit{bidirectional} representations that both serve as an output modality to interpret data and an \textit{input} modality to proactively author concepts and investigate new research questions.

\section{LLooM: Concept Induction using Large Language Models}
We define \textit{concept induction} as a process that takes an unstructured text dataset as input and produces a set of emergent, high-level concepts as output, each of which are defined by explicit criteria. 
We first describe LLooM, a concept induction algorithm that leverages large language models to iteratively extract and synthesize concepts from raw data.
Then, we present the LLooM Workbench, a text analysis tool that uses the LLooM algorithm to enable analysts to generate, visualize, and refine high-level concepts from text data.

\subsection{The LLooM Algorithm}
\label{section:lloom_alg}

\begin{figure*}[!tb]
  \includegraphics[width=\textwidth]{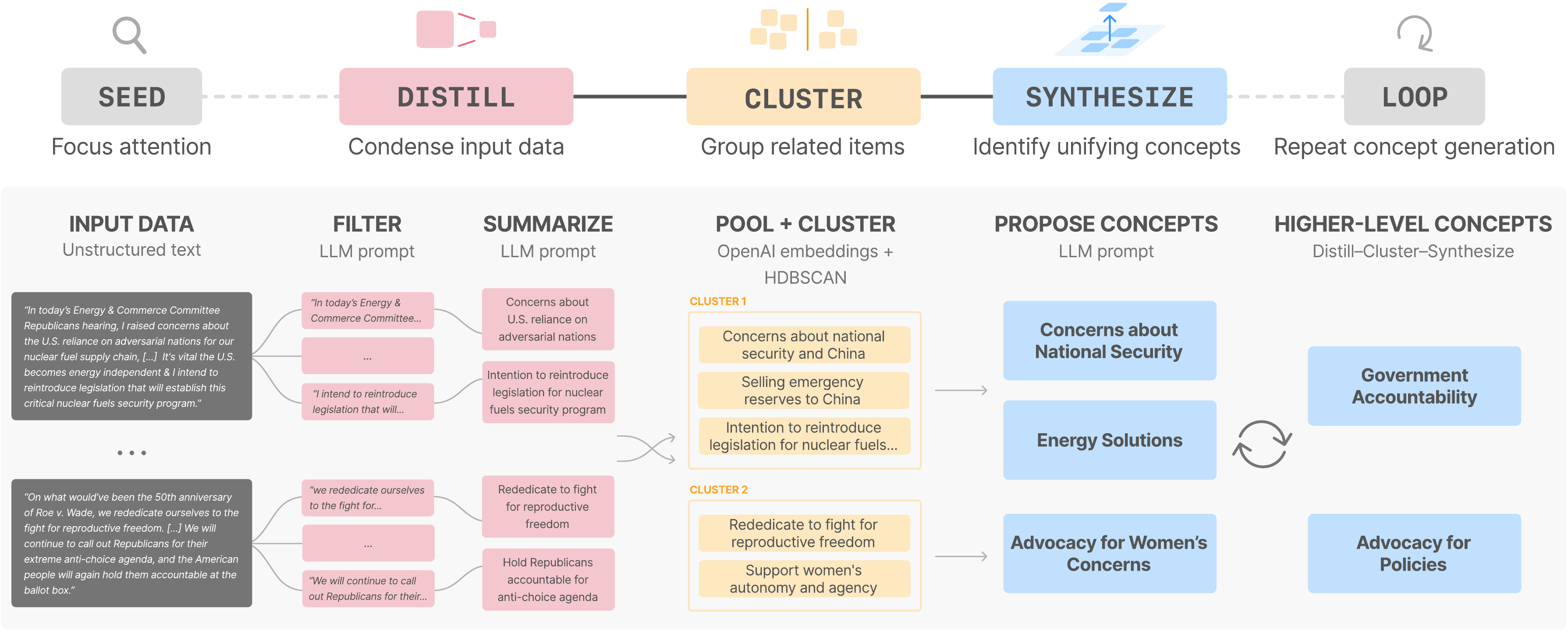}
  \caption{
    \textbf{Concept generation in the LLooM algorithm}, demonstrated with sample text inputs. The process starts with unstructured text data and an optional \texttt{Seed} from the analyst. Then, the \texttt{Distill} operator condenses the input data with an LLM by filtering to excerpts and summarizing to bullet points (\texttt{gpt-3.5-turbo}). The \texttt{Cluster} operator pools and groups the distilled bullet points using a clustering algorithm. Finally, the \texttt{Synthesize} operator proposes high-level concepts using an LLM prompt (\texttt{gpt-4}). The \texttt{Loop} operator can optionally repeat this process multiple times to produce higher-level concepts.
  }
  \label{fig:lloom_alg}
  \Description{The figure starts on the left with a Seed operator block above example blocks of input text data. This leads to a section labeled by the Distill operator above a Filter step that extracts excerpts from the original text and a Summarize step that provides bullet points for the excerpts. The next section is labeled by the Cluster operator above a set of bullet point blocks that have been grouped into clusters. The next section shows the Synthesis operator above concept blocks with names like ``Concerns about National Security.'' Finally, there is a circular arrow indicating iterative rounds before a Loop operator section that shows higher-level concepts like ``Government Accountability.''}
\end{figure*}

The LLooM algorithm performs concept induction by executing iterative rounds of concept \textit{generation} and \textit{scoring} using a large language model (LLM). We specifically use GPT-3.5 and GPT-4 in our implementation.
Summarized in Figure~\ref{fig:lloom_alg}, the primary goal of our algorithm is to execute the critical \textit{synthesis} step of bridging from low-level text signals to high-level concepts, which we define as human-interpretable descriptions defined by explicit \textit{inclusion criteria}, specifically a natural-language description of decision rule(s) for whether an input matches the concept. With prior methods, analysts must carry out this critical bridging work from low-level text signals to high-level concepts themselves; LLMs provide assistance with this step. 

First, for the \textit{concept generation} step, LLooM implements the \badge[bl]{Synthesize} operator that prompts the LLM to generalize from provided examples to generate concept descriptions and criteria in natural language. 
As we demonstrate empirically in our technical evaluations (\S\ref{section:tech_eval}), directly prompting an LLM like GPT-4 to perform this kind of synthesis produces broad, generic concepts rather than nuanced and specific conceptual connections (e.g., that a set of posts are \textit{feminist-related}, rather than that they all constitute \textit{men's critiques of feminism}). While generic concepts may be helpful for an overarching summary of data, analysts seek richer, more specific concepts that characterize \textit{nuanced patterns} in the data, as supported by our expert case studies (\S\ref{section:eval}). Additionally, such synthesis is not possible for text datasets that exceed LLM context windows.

To address these issues, the LLooM algorithm includes two operators that aid both data size and concept quality: (1)~a \badge[rd]{Distill} operator, which shards out and scales down data to the context window while preserving salient details, and (2)~a \badge[yl]{Cluster} operator, which recombines these shards into groupings that share enough meaningful overlap to induce meaningful rather than surface-level concepts from the LLM.

Finally, for the \textit{concept scoring} step, we leverage the zero-shot reasoning abilities of LLMs to implement a \badge[gr]{Score} operator that labels data examples by applying concept criteria expressed as zero-shot prompts. With these labels, we can visualize the full dataset in terms of the generated concepts or further iterate on concepts by looping back to concept generation. We now walk through the LLooM algorithm in detail.

\subsubsection{Concept Generation}
The key to our concept induction algorithm is the \texttt{Synthesize} operator, which leverages the capability of LLMs to synthesize high-level, conceptual similarities shared among sets of examples. When chained together with other auxiliary operators to form a {\texttt{Distill}--\texttt{Cluster}--\texttt{Synthesize}} pipeline, the \texttt{Synthesize} operator allows the LLooM algorithm to generate high-level concepts ~(Figure~\ref{fig:lloom_alg}).

\paragraph{\badge[bl]{Synthesize}}
This operator takes as input a group of text examples and is tasked with producing one or more unifying, \textit{high-level concepts} that connect the examples. By our definition, these high-level concepts must consist of both a human-understandable description and inclusion criteria. 
LLMs have capabilities that are well-suited to aid this task. For example, GPT-3.5 Turbo and GPT-4 can successfully generalize from a small number of examples; i.e., to identify unifying concepts and carry them forward to new examples. This capability, also referred to as few-shot reasoning, is often leveraged in cases where the user \textit{already knows} the underlying pattern and wants the model to apply it repeatedly (e.g., to translate text to different formats, or to transfer a writing style)~\cite{brown2020language}. However, we can also leverage this capability in situations where the user \textit{does not know} ahead of time what concepts exist in their data to aid discovery. While LLMs can hallucinate and produce unreliable outputs, by constructing our task to not just produce concepts, but the criteria to evaluate those concepts, we can verify LLM outputs by reviewing the criteria and re-evaluating the original data to test if concepts hold.

Building on this insight, LLooM implements the \texttt{Synthesize} operator as a zero-shot prompt that instructs an LLM (\texttt{gpt-4}) to identify unifying high-level concepts from a provided cluster of examples. The instructions ask the model to generate a \textit{name} that describes the concept, provide IDs of the \textit{representative examples} that best match this concept, and generate its own \textit{prompt} that can evaluate a novel text example and determine whether the concept applies. Each of these components is useful output for understanding the meaning of a concept. These components also leverage a chain-of-thought (CoT) prompting strategy~\cite{wei2022chain,kojima2022large} that instructs the model to provide a trace of its work and improve the likelihood of internal consistency.

We include our prompt template below.\footnote{Within the prompt, we use the term ``pattern'' as a synonym for ``concept''; through experimentation, we found that this term was more effective for concisely conveying that the concepts needed to be shared among multiple items, while ``concept'' is a more generic term that resulted in less reliable instruction-following.}  Users can vary the concept name length, the number of representative concept examples, and the number of concepts to suggest; we use 2-4 word concept names and request 1-2 representative examples by default. 
\begin{lstlisting}[language=Markdown]
    I have this set of bullet point summaries of text examples:
    {bullets_json}
    
    Please write a summary of {n_concepts} unifying patterns for these examples {seed_phrase}.
    For each high-level pattern, write a {n_name_words} word NAME for the pattern and an associated one-sentence ChatGPT PROMPT that could take in a new text
    example and determine whether the relevant pattern applies. 
    Please also include {n_example_ids} example_ids for items that BEST exemplify the pattern. Please respond ONLY with a valid JSON in the following format:
    {{
        "patterns": [ 
            {{
                "name": "<PATTERN_NAME_1>", 
                "prompt": "<PATTERN_PROMPT_1>", 
                "example_ids": ["<EXAMPLE_ID_1>", "<EXAMPLE_ID_2>"]
            }},
            {{
                "name": "<PATTERN_NAME_2>", 
                "prompt": "<PATTERN_PROMPT_2>", 
                "example_ids": ["<EXAMPLE_ID_1>", "<EXAMPLE_ID_2>"]
            }},
        ]
    }}
\end{lstlisting}
Notably, this operator starts where topic modeling typically ends: with data groupings that are likely to share similarities. However, in contrast to approaches that seek to assign a label to clusters, a key differentiator of our \texttt{Synthesis} operator is that it is not bound to labeling an entire group of examples, but frames the task around \textit{selectively} proposing salient connections among items in a group. Our prompt instantiates this by asking the model to identify subsets of examples that best exemplify concepts rather than requiring that all examples match the concept and phrasing the task as pattern identification rather than holistic label assignment.
Since clusters are often noisy, instead of attempting to holistically summarize the cluster, which could lead to a vague connection, our approach is to identify pockets of examples that have unifying connections.

\paragraph{Auxiliary operators} 
The remaining operators of the concept generation phase are designed to improve the performance of our core \texttt{Synthesize} operator by mitigating several challenges of large language models, such as token limits and uneven output quality.

\badge[rd]{Distill}.
The \texttt{Distill} operator condenses input data into a more compact representation while preserving important or distinctive attributes, which both addresses LLM context window limits and grants the ability to ``zoom'' into areas of interest to improve concept generation.
In LLooM, we take a multi-step approach to implement our \texttt{Distill} operator in natural language. First, we perform a Filter step of zero-shot summarization by providing the input text example and prompting an LLM (\texttt{gpt-3.5-turbo}) to generate an \textit{extractive} summarization that selects exact quotes from the original text; this step can be omitted if the text is not very long. Users can adjust the number of quotes to select, but by default the parameter is left empty such that the model may extract any number of quotes. Below is an example of the Filter prompt:
\begin{lstlisting}[language=Markdown]
    I have the following TEXT EXAMPLE:
    {text_example_json}
    
    Please extract {n_quotes} QUOTES exactly copied from this EXAMPLE {seed_phrase}. 
    Please respond ONLY with a valid JSON in the following format:
    {{
        "relevant_quotes": [ "<QUOTE_1>", "<QUOTE_2>", ... ]
    }}
\end{lstlisting}

Then, we perform a Summarize step, which prompts an LLM (\texttt{gpt-3.5-turbo}) to generate an \textit{abstractive} summarization in the form of bullet point text summaries. Users can adjust the number of bullet points to generate and the length of the bullet points if necessary, but we use a default of ``2-4'' bullet points with lengths of ``5-8'' words. We include an example prompt below:
\begin{lstlisting}[language=Markdown]
    I have the following TEXT EXAMPLE:
    {text_example_json}
    
    Please summarize the main point of this EXAMPLE {seed_phrase} into {n_bullets} bullet points, where each bullet point is a {n_words} word phrase. 
    Please respond ONLY with a valid JSON in the following format:
    {{
        "bullets": [ "<BULLET_1>", "<BULLET_2>", ... ]
    }}
\end{lstlisting}

The \texttt{Distill} operator allows us to pare down each example to its salient attributes and is inspired by initial line-by-line coding or open coding in qualitative analysis~\cite{muller2014curiosity, charmaz2006constructing}.

\badge[yl]{Cluster}.
Next, the \texttt{Cluster} operator groups together related items based on patterns in their representations from the \texttt{Distill} step. 
For the \texttt{Cluster} operator to generate \textit{cross-cutting} concepts, all of the distilled bullet points are detached from their original examples and pooled together. Thus, the input of the \texttt{Cluster} operator is the set of condensed bullet points from the \texttt{Distill} operator, and the output is a set of group assignments, such that each isolated bullet point is assigned to a group of related items.
The LLooM algorithm transforms bullet points into embeddings using a specified pre-trained embedding model and then clusters the items using a provided clustering algorithm.
Our implementation uses OpenAI's \texttt{text-embedding-ada-002} model due to its relatively long context and fast generation time. For clustering,  we select HDBSCAN, a hierarchical clustering algorithm, because its density-based approach does not require heavy parameter tuning and does not require all points to be placed in a cluster. These properties increase the likelihood that our dynamically-generated clusters will contain salient examples without manual intervention.
The \texttt{Cluster} operator resembles the initial phases of processes like affinity grouping and axial coding in that it coalesces examples into possible groupings, which is a critical step before the \texttt{Synthesize} operator can complete the process to identify similarities and conceptual themes.

\badge[gr]{Seed}. What if the analyst wants to steer LLooM's attention toward particular aspects of the data? LLooM allows the analyst to guide the system to attend to ``social issues'' for a political dataset, ``evaluation methods'' for an academic papers dataset, or ``displays of emotion'' for a text conversations dataset. The optional \texttt{Seed} operator accepts a user-provided \textit{seed term} to condition the \texttt{Distill} or \texttt{Synthesize} operators, which can improve the quality and alignment of the output concepts. This seed term provides additional instructions in the LLM prompt to ask the model to attend to a particular aspect of the data.\footnote{The seed term is inserted as the \texttt{seed\_phrase} shown in the example prompts above in the format ``related to \texttt{\{seed\_term\}}.''} For the \texttt{Distill} operator, this will instruct the model to generate summaries that focus on parts of the data related to the seed term. Similarly, for the \texttt{Synthesize} operator, this will instruct the model to propose unifying concepts among the examples that are related to the seed term.
Taking inspiration from qualitative analysis, which acknowledges that there are multiple valid interpretations of data, the \texttt{Seed} operator grants the analyst control to steer the concept generation process based on their analysis goals and desired interpretive lens.

\subsubsection{Concept Scoring}
The concept generation phases of the LLooM algorithm are followed by a concept scoring phase that applies the generated concepts back to the full dataset. 

\paragraph{\badge[gr]{Score}}
Armed with the concepts, LLooM next applies a score (e.g., 0-1) that describes the association between each input and the concept. For each high-level concept, the system applies the \texttt{Score} operator to all examples (input texts) to generate a concept score that estimates how well each example matches the generated concept prompt. This is implemented using a batched zero-shot prompt that includes a set of examples in JSON format, the concept prompt, and instructions to generate an answer in multiple-choice format. 
Prior work has found that LLMs do not provide calibrated 0-1 confidence scores in zero-shot settings~\cite{lin2022teaching}. However, recent work has found that for instruction-tuned OpenAI models such as GPT-3.5, multiple choice prompting~\cite{santurkar2023opinions,robinson2022leveraging} can provide approximate answer probabilities. 
We use multiple choice prompting to instruct the model to generate a multiple-choice answer\footnote{Our multiple choice options are: A: Strongly agree, B: Agree, C: Neither agree nor disagree, D: Disagree, E: Strongly disagree} for each provided example along with a rationale. These answers are parsed and converted to bucketed numerical scores with ``Strongly agree'' mapping to 1.0 and ``Strongly disagree'' mapping to 0.0. The scores are then thresholded to a binary label; users can adjust the threshold at which an example should be considered a concept match. Given $n$ examples and $c$ high-level concepts, this phase results in a $n\times c$ matrix with a binary concept label for each example.

This concept scoring phase is designed to bring some of the benefits of the \textit{deductive coding} process in qualitative analysis, which applies codes back to the data. This deductive coding process both allows an analyst to make sense of their data and also exposes potential gaps, biases, or limitations in their codebook, which can be addressed in further iterations of inductive coding.

\paragraph{\badge[gr]{Loop}}
Finally, based on the concept scoring results, LLooM can use a \texttt{Loop} operator to execute multiple iterations of the algorithm. 
This operator executes the logic to \textit{revise the inputs} to the next iteration of the pipeline. 
We use \textit{data coverage} to determine which examples will be processed in each subsequent iteration. After the concept scoring phase completes, the \texttt{Loop} operator identifies two classes of outliers: 1) \textit{not-covered} examples, which did not match \textit{any} of the current high-level concepts and 2) \textit{covered-by-generic} examples, which only matched ``generic'' concepts, those that matched a majority of examples (at least $50\%$). All such examples are provided as input to the next iteration of the algorithm, and the concepts generated by subsequent runs are added to the full set of concepts.

\subsubsection{Implementation Details}
The LLooM algorithm is implemented as a Python library that can be imported into computational notebooks like Jupyter or web application frameworks like Flask. 
We primarily use GPT-3.5 (\texttt{gpt-3.5-turbo}) for all operators except for the \texttt{Synthesize} operator, which benefits from the improved reasoning capabilities of GPT-4.
For the \texttt{Distill} operator, both the Filter and Summarize steps are executed with zero-shot prompts to the \texttt{gpt-3.5-turbo} model using the OpenAI API with a temperature of $0$ to provide more consistent results. 
For the \texttt{Cluster} operator, we use OpenAI embeddings from the \texttt{text-embedding-ada-002} model, and we use the HDBSCAN clustering algorithm. 
For the \texttt{Synthesize} operator, we use the OpenAI API with options for either \texttt{gpt-3.5-turbo} or \texttt{gpt-4}, again using a temperature of $0$.
The \texttt{Score} operator provides options to use either the OpenAI API with \texttt{gpt-3.5-turbo} or the Google PaLM API with the \texttt{chat-bison-001} model, both with a temperature of $0$ for consistency.
As a point of reference, across the scenarios that we describe in \S\ref{section:lloom_scenarios}, the total cost of one run of the LLooM algorithm averaged \$1.44 in total cost, used $848,323$ tokens (combining input and output), and took on average $13.7$ minutes to complete. Notably, the \textit{concept scoring} step is substantially more costly and time-intensive than the \textit{concept generation} step, on average consuming $79.9\%$ of the total cost and $58.4\%$ of the total time.
Full prompts are provided in Appendix~\ref{appendix:prompts}.

\begin{figure*}[!tb]
  \includegraphics[width=1.0\textwidth]{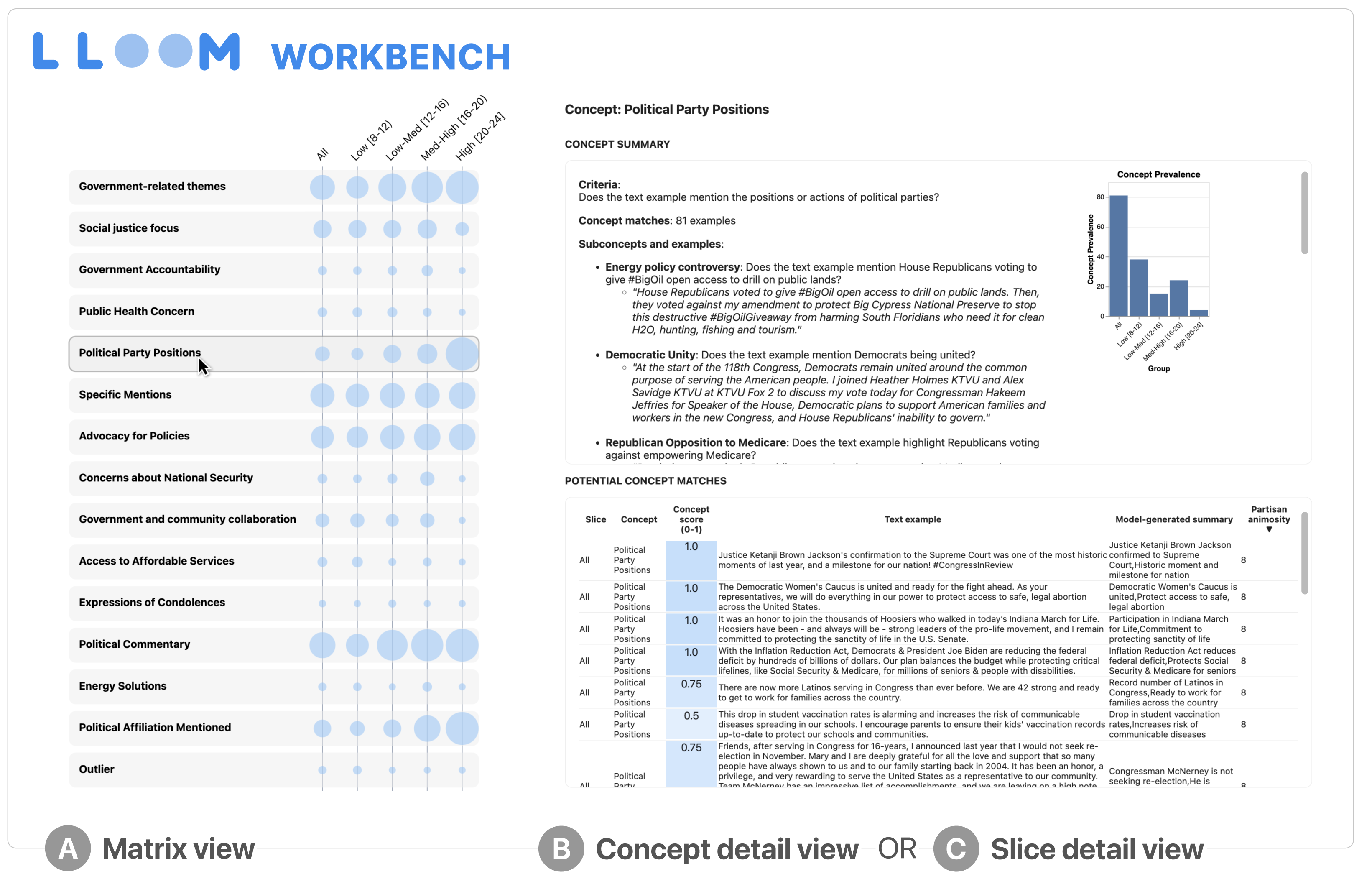}
  \caption{
    \textbf{The LLooM Workbench}, an interactive text analysis tool that leverages LLooM's concept induction capabilities. The tool consists of the (A)~Matrix view with an overview of the prevalence of concepts among user-defined data slices. Selecting a concept row displays the associated (B)~Concept detail view, which displays the concept criteria, subconcepts, and matching examples. Selecting a slice column displays the corresponding (C)~Slice detail view, which displays a similar overview of the examples within the slice.
  }
  \label{fig:visualization}
  \Description{The figure displays a screenshot of the LLooM Workbench interface. The Matrix view on the left has a grid of concept rows (e.g., ``Government-related themes,'' ``Social justice focus'') and slice columns (e.g., ``All'', ``Low (8, 12]''). The cells contain circles whose size is based on prevalence. The right upper half has a card listing the concept name, criteria, subconcepts, and a histogram of concept prevalence across slices. The right bottom half has a scrollable table view with text columns, concept scores, and metadata columns.}
\end{figure*}

\subsubsection{Algorithm Limitations}
We note several limitations of the current LLooM algorithm that may be fruitful areas for future work. 
First, the LLooM algorithm has a number of available parameters, such as the number of quotes to extract and the number of bullet points to generate in the \texttt{Distill} phase. While these parameters are interpretable to a user, they are not straightforward for a user to set in advance, so it would be best for the system to dynamically set these values when possible. Our system has default values and formulas to calculate parameter values, but these have not been robustly tested for appropriateness on a wide variety of datasets.

Additionally, the current implementation does not make use of verification steps, for example to ensure that quotes are exact matches, that bullet points are accurate to quotes, and that concept scores and rationale appear correct. While reliable verification is an ongoing challenge for LLMs, future extensions of LLooM could benefit from programmatic checks and LLM operators explicitly designed to verify outputs at each phase.
Our use of LLMs also means that there is variability in the results upon re-run. While this can be a useful feature to explore parallel analysis paths and simulate variations, it may be undesirable in cases where analyses must be replicable or where robust, consistent alignment is necessary~\cite{chuang-etal-2015-topiccheck}.

\subsection{The LLooM Workbench}
\label{section:lloom_workbench}
We instantiate the LLooM concept induction algorithm in an interactive text analysis tool called the LLooM Workbench. With this tool, an analyst can upload their unstructured text dataset, and LLooM will automatically extract and display concepts in an interactive visualization (Figure~\ref{fig:visualization}).

\subsubsection{Workbench Components}
The LLooM Workbench allows analysts to see and interact with data in terms of high-level concepts.

\textit{Matrix View}. Concept threads are the focal point of the workbench's matrix visualization (Figure~\ref{fig:visualization}A). In this view, the generated concepts are displayed as rows, and user-specified data slices are displayed as columns. By default, an ``All'' slice is initially shown for all datasets, but users can specify their own custom slices by authoring filters on any metadata column from the original dataset or any generated concept. 
Then, each cell in the matrix at the intersection of concept $c$ and slice $s$ displays a circle whose size indicates the prevalence of concept $c$ in slice $s$, and can be normalized by the total size of the concept or the total size of the slice. This visualization allows users to perform consistent comparisons of a particular concept's prevalence across data slices (within a row) or of all concepts' prevalence within a particular slice (within a column).
The user can select any row to dive into a Concept Detail View or a column to dive into a Slice Detail View.

\textit{Concept Detail View}. In this panel, a user can both inspect the meaning of a selected concept and review the subset of the dataset that matched this concept (Figure~\ref{fig:visualization}B). The upper left portion of the panel displays a concept summary that includes the generated concept name, the generated criteria (which is executed to evaluate whether unseen examples match the concept), subconcepts that led to this concept, and representative text examples for each subconcept. The upper right side of the panel displays a histogram for a more detailed view of concept prevalence across slices. Finally, the bottom section of the panel displays a concept match table, which displays examples that potentially match the concept based on LLooM concept scores. The primary dataset text column and concept score column are displayed by default, but users can specify to include any additional column from the original dataset. For cases where the algorithm performed the Filter step to extract relevant quotes, the filtered text is highlighted in the table.

\textit{Slice Detail View}. Similarly, this panel displays details of a user-defined slice. The upper portion of the panel displays the user-provided slice name (e.g., ``Low toxicity'') and filtering criteria (e.g., \texttt{toxicity < 0.25}), along with a histogram for a more comprehensive view of concept prevalence for the slice (Figure~\ref{fig:visualization}C). The bottom of the panel displays a slice summary table, which includes all examples that met the filtering criteria. Each row in the table represents an example, and the table displays the primary text column and all concept score columns by default; users can again specify to include any additional metadata column from the dataset.

\subsubsection{Workbench Actions}
In addition to the core visualizations, the LLooM Workbench supports a range of actions for analysts to build on the initial set of LLooM concepts.

\textit{Adding and Editing}. Users can manually add custom concepts by specifying a concept name and an associated criteria prompt that defines the concept. The concept will be applied to the data with the \texttt{Score} operator, and it will be added to the matrix visualization as an additional row. Users may also edit an existing concept by modifying its name and/or criteria prompt, and they can similarly initiate concept rescoring after making these modifications.

\textit{Merging and Splitting}. 
Users can also merge multiple related concepts, which prompts the system to generate a new concept name and criteria that combine the selected concepts. Conversely, users can split concepts when they are too general, which prompts the system to author new subconcepts for the selected concept.

\subsubsection{Implementation Details}
The LLooM Workbench is implemented as Jupyter widget for use in computational notebooks. The widget draws on the LLooM algorithm Python library described in \S\ref{section:lloom_alg} and implements a library of Svelte UI components. We use the anywidget Python library\footnote{\url{https://anywidget.dev}} to render the Svelte components as notebook widgets. The interactive LLooM matrix visualization is implemented using the D3 JavaScript library.\footnote{\url{https://d3js.org}}

\section{LLooM Scenarios}
\label{section:lloom_scenarios}

By surfacing conceptual threads as an interpretable and malleable material with which to work with data, LLooM opens up new ways to understand and interact with text data. 
In the next three sections, we walk through a multi-part evaluation to: demonstrate the concepts that LLooM surfaces from a variety of real-world datasets (\S\ref{section:lloom_scenarios}: LLooM Scenarios), understand the technical performance of the LLooM algorithm compared to existing approaches (\S\ref{section:tech_eval}: Technical Evaluations), and explore how expert analysts make sense of data with concepts in the LLooM Workbench (\S\ref{section:eval}: Expert Case Studies).

First, to demonstrate LLooM's outputs on real-world datasets in a variety of domains, we present four data analysis scenarios: developing content moderation policies for toxic content (\S\ref{section:case_study_tox_fem}), mitigating partisan animosity on social media (\S\ref{section:dem_attitudes}), analyzing academic paper abstracts (\S\ref{section:hci_papers}), and investigating anticipated consequences of AI research (\S\ref{appendix:case_study_neurips}).
These cases were selected to span a variety of text formats and lengths (from short social media posts to paper abstracts) and analysis goals (from surveying literature to developing a decision-making policy or ML model).  

\subsection{Method}
The goal of the scenarios is to qualitatively illustrate how LLooM works in practice. Thus, we compare against topic models because they are the de facto standard in unstructured text analysis today. 

\subsubsection{Baseline result generation}
We use a state-of-the-art BERTopic model as a representative baseline topic model. For each scenario, we ran BERTopic using OpenAI \texttt{text-embedding-ada-002} embeddings and HDBSCAN with a minimum cluster size set to $2-3\%$ of the full dataset size. Then, we gathered all resulting topics and their associated keywords (generated by BERTopic using c-TF-IDF) along with the documents assigned to each topic.
To run LLooM, we initiated a new session that executed one iteration of the LLooM process. 
Within LLooM, we randomly sampled up to 200 items to run this process and set a limit of at most 20 final concepts to generate. We focused on data samples of these sizes to prioritize \textit{interactive} concept induction completion times ranging from 5-15 minutes and concept scoring times under 20 seconds to support manual concept authoring.
For these runs, we used \texttt{gpt-3.5-turbo} to perform all distilling and synthesizing operations, and we used OpenAI \texttt{text-embedding-ada-002} embeddings for the clustering phase. To assign items to concepts, we gathered all items that received a positive label for each concept, using a threshold set at the highest score option (1.0: Strongly agree).

\begin{figure*}[!tb]
  \includegraphics[width=1.0\textwidth]{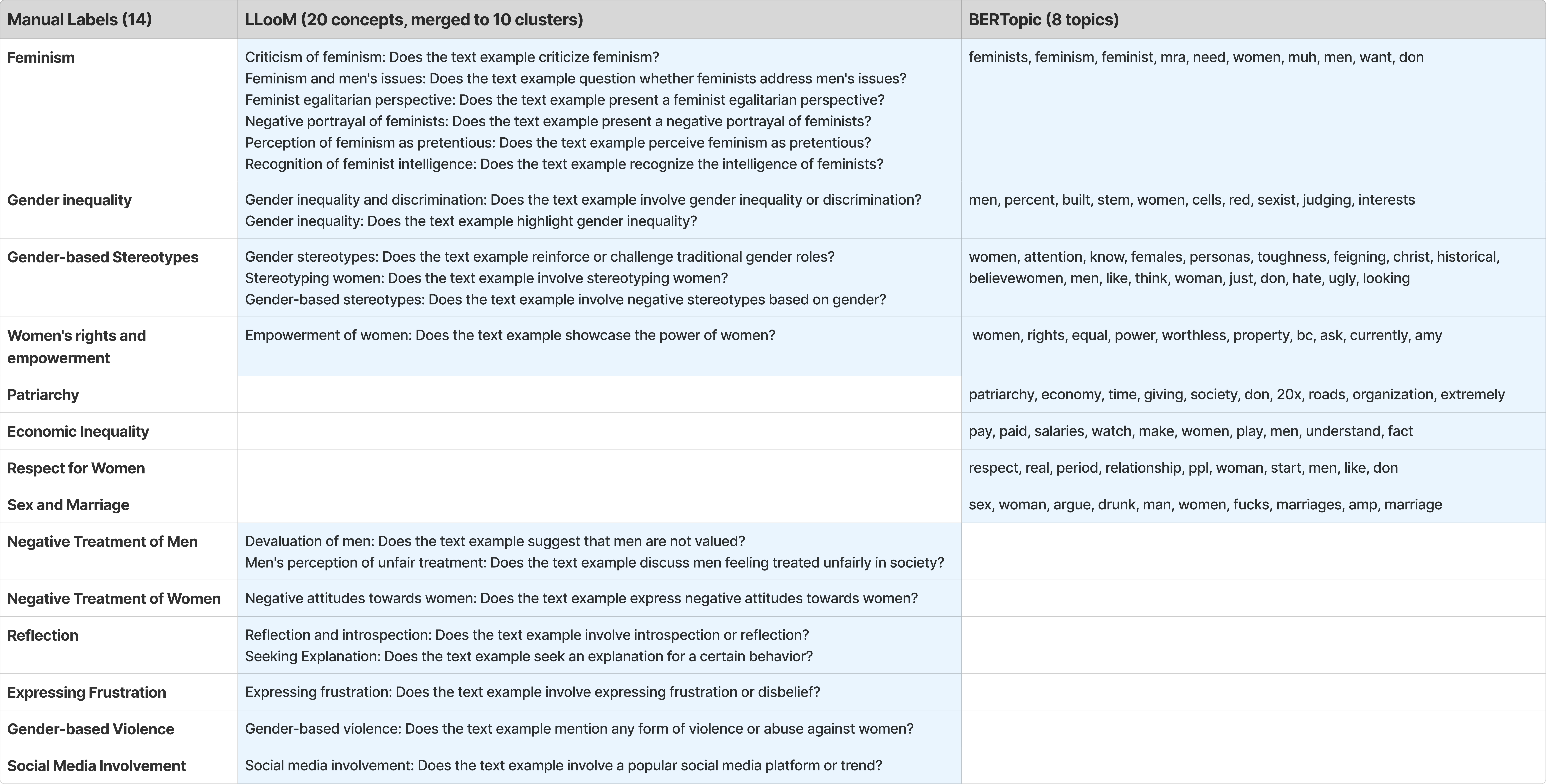}
  \caption{
    For the toxic content dataset, LLooM generates content-related concepts such as \textit{Empowerment of women} and \textit{Gender inequality}, but also surfaces style- and tone-related concepts such as \textit{Expressing frustration} and \textit{Reflection and introspection}.
  }
  \label{fig:baseline_tox_fem}
  \Description{The figure displays a three-column table with Manual Labels, LLooM concepts, and BERTopic topics. Each column contains cells with the corresponding output type, and each row contains any overlapping results from across the three methods. This table includes rows for feminism, gender inequality, gender-based stereotypes, women’s rights and empowerment, patriarchy, economic inequality, respect for women, sex and marriage, negative treatment of men, negative treatment of women, reflection, expressing frustration, gender-based violence, and social media involvement.}
\end{figure*}

\subsubsection{Baseline qualitative analysis}
For each dataset, a member of the research team manually reviewed all results. For BERTopic, they reviewed each topic by inspecting the generated keywords (e.g., ``oil, gas, energy,'' ``house, republicans, democrats'') and all documents assigned to the topic, and they wrote their own manual label to synthesize the unifying theme of the topic (e.g., \textit{Environmental policy}, \textit{Political parties}). 

By design, LLooM has the advantage of generating highly specific concepts described in natural language (e.g.,  \textit{User interface enhancement} and \textit{User experience enhancement}). However, BERTopic outputs are unlikely to communicate such nuance with keywords alone (e.g., ``user, users, interaction''), so it would seem unfair to penalize the method largely because it lacks such expressivity.
Thus, to facilitate a direct comparison with BERTopic outputs, we take a conservative approach to estimate overlap by grouping together sets of LLooM concepts that would be unreasonable for BERTopic to produce.
The research team member reviewed all LLooM concepts and grouped together any concepts that overlapped in meaning: either if one concept was a subset of another concept (e.g., \textit{Advocacy for Policies} and \textit{Advocacy}), or if two concepts appeared to be synonymous (e.g., \textit{User interface enhancement} and \textit{User experience enhancement}). 
Using this simplified set of results, BERTopic topics and LLooM concepts deemed as having shared meaning were considered \textit{overlapping} results.

\subsection{Scenario 1: Developing Moderation Policies for Toxic Content}
\label{section:case_study_tox_fem}
First, we investigate a \textit{content moderation} task where a social media platform is developing a model to perform automated content moderation of text posts. 
Prior research has found substantial disagreement among the population on what constitutes toxic content~\cite{kumar2021designing, disagreement_deconv}, so unstructured text analysis might grant moderators greater nuance in understanding and triaging emergent user behavior. 
We use a dataset of social media posts (from Twitter, Reddit, and 4chan) that gathers a diverse set of annotators' perspectives on content toxicity with ratings from $17,280$ U.S. survey participants on over $100,000$ examples~\cite{kumar2021designing}. 
We applied BERTopic to the full dataset, filtered to the largest clusters, and selected the feminism-related cluster ($n=496)$ because it aligned with a distinct user community and potentially controversial topics.

\subsubsection{Results}
\begin{figure*}[!tb]
  \includegraphics[width=0.9\textwidth]{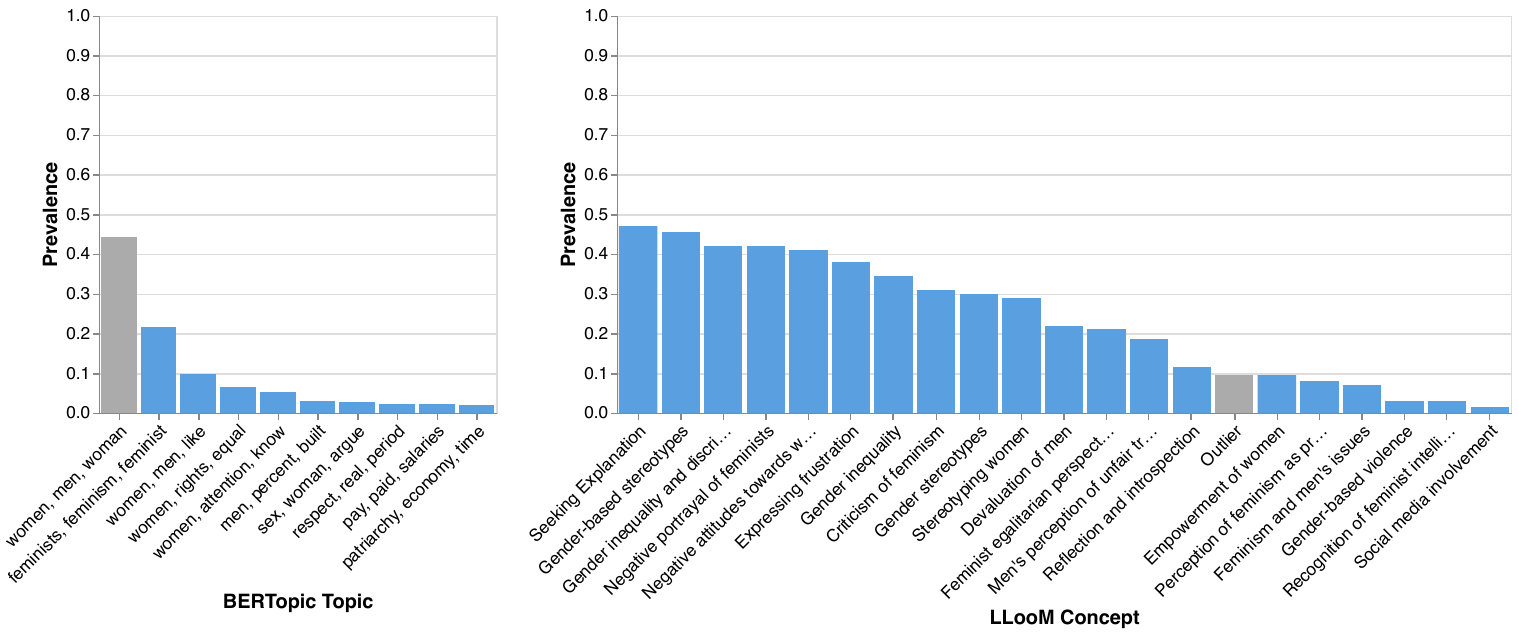}
  \caption{
   Scenario 1: Toxic content dataset---BERTopic places a large proportion of examples ($44.3\%$) in an uncategorized cluster (in grey) while most other clusters contain between $2-10\%$ of examples. LLooM concepts display a range of prevalence values from $1-50\%$, and the outlier category contains $9.5\%$ of examples.
  }
  \label{fig:hist_tox_fem}
  \Description{This figure displays two histograms, with BERTopic on the left and LLooM on the right. The x-axis lists topics or concepts, respectively, and the y-axis has a 0-1 range and displays the prevalence of concepts among the examples. BERTopic has one large uncategorized cluster with 44\% of examples, and all other clusters have around 20\% or lower. LLooM has 12 concepts that contain over 20\% of the examples, and its outlier concept has 9.5\% of examples.}
\end{figure*}

LLooM generated $10$ unique sets of concepts, such as ``Devaluation of men,'' ``Empowerment of women,'' and ``Gender inequality and discrimination,'' as summarized in Figures~\ref{fig:baseline_tox_fem} and \ref{fig:hist_tox_fem}. Meanwhile, BERTopic generated $8$ topics with keywords such as ``feminists, feminism, feminist'' and ``women, men, like.'' 
Based on manual inspection of the BERTopic results, these were fairly high-level groupings aligned with particular keywords such as feminism, power, and men/women. Meanwhile, LLooM results were not bound to keywords, but often captured attitudes (e.g., ``Devaluation of men'') and interpretations (e.g., ``Men's perception of unfair treatment,'' ``Reflection and introspection'') that went beyond surface-level features of text. 
We observed that $50\%$ of BERTopic results were covered by LLooM while $40\%$ of LLooM results were covered by BERTopic, so there was some divergence between the two methods.
In addition, $44.4\%$ of examples were uncategorized by BERTopic, while $9.5\%$ were uncategorized by LLooM, so LLooM achieved higher data coverage.

\subsection{Scenario 2: Mitigating Partisan Animosity on Social Media}
\label{section:dem_attitudes}
Political polarization is a dominant concern in the United States, and it poses potential existential risks to democracy. If social media algorithms play a role in amplifying partisan animosity~\cite{milli2023twitter,jia2023embedding}, how might we redesign social media algorithms to mitigate this effect? Our next scenario investigates political social media posts to explore whether we can detect and downrank content that amplifies partisan animosity.
We use a dataset of public Facebook posts from~\citet{jia2023embedding}.
This dataset was generated by filtering for political posts on CrowdTangle using politics-related page categories such as ``politics,'' ``politician,'' ``political organization,'' and ``political party.'' 
The dataset consists of $405$ posts that were randomly sampled and manually coded for partisan animosity.\footnote{The scores consist of 8 sub-scores that are summed together. Each sub-score can range from 1-3, so the score range is from 8 to 24, where 8 corresponds to the lowest partisan animosity and 24 corresponds to the highest partisan animosity.}

\begin{figure*}[!tb]
  \includegraphics[width=0.8\textwidth]{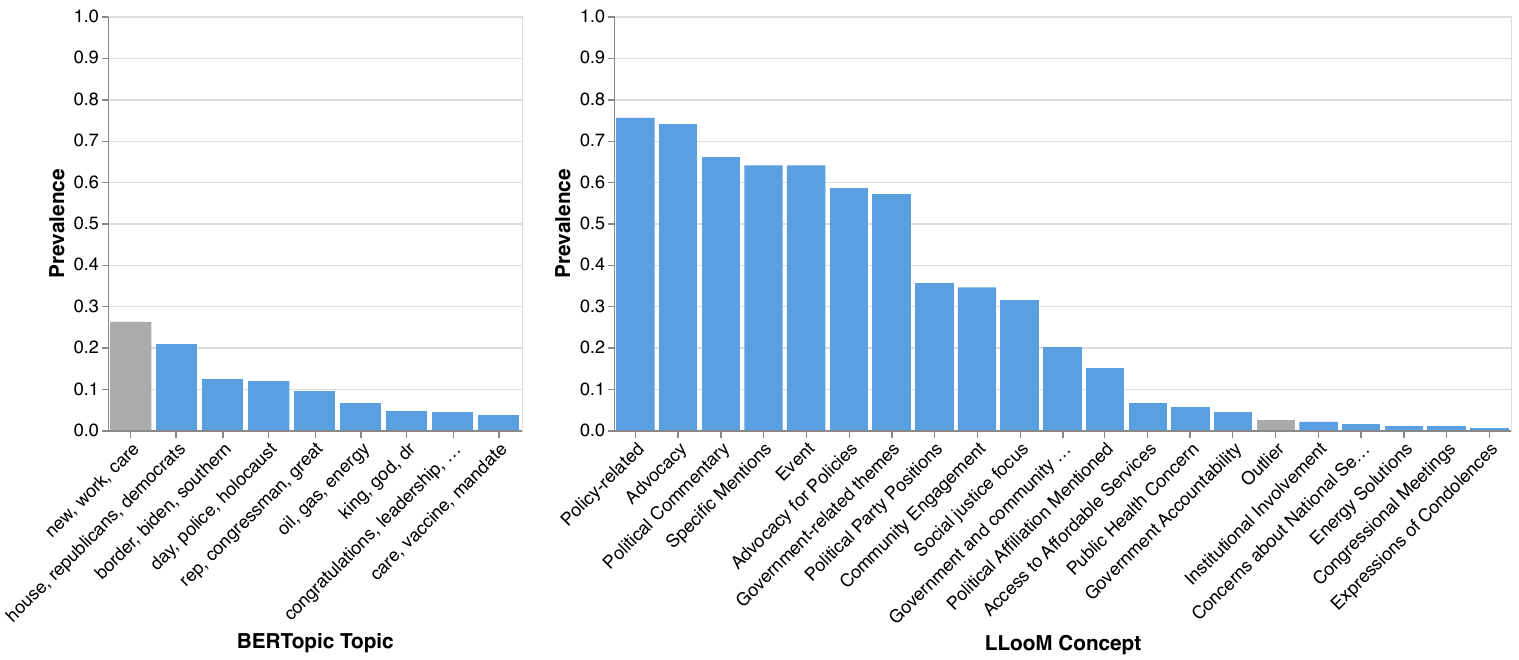}
  \caption{
    Scenario 2: Partisan animosity dataset---The largest topic from BERTopic is again an uncategorized topic with $26.2\%$ of examples. A set of seven LLooM concepts captured more generic, high-prevalence political topics, but there is a range of concept prevalence values, and only $2.5\%$ of examples were outliers.
  }
  \label{fig:hist_dem_feeds}
  \Description{(Same histograms format as Figure~\ref{fig:hist_tox_fem}) BERTopic has one larger uncategorized cluster with 26.2\% of examples, and all clusters have 20\% or lower. LLooM has 11 concepts that contain more than 20\% of examples, and its outlier concept has 2.5\% of the examples.}
\end{figure*}

\subsubsection{Results}
LLooM generated $14$ distinct concepts, such as ``Concerns about National Security,'' ``Political Affiliation Mentioned,'' and ``Advocacy for Policies,'' summarized in Figure~\ref{fig:hist_dem_feeds}.
Meanwhile, BERTopic generated $8$ topics with keywords such as ``house, republicans, democrats,'' ``care, vaccine, mandate,'' and ``oil, gas, energy.''  
BERTopic produced data groupings that aligned with major entities (e.g., manual labels of ``Political Parties'' and ``Community'') and political issues (e.g., manual labels of ``Border Policy'' and ``Environmental Policy'').
LLooM concepts similarly covered many of these same entities and political issues, but also captured certain \textit{user behaviors} such as expressions of condolences and specific mentions of individuals (such as political figures) in the Facebook posts. LLooM also captured several additional political issues such as social justice and access to affordable services.
While $87.5\%$ of BERTopic results were covered by LLooM, $50\%$ of LLooM results were covered by BERTopic, so there was a sizeable portion of LLooM concepts that were novel additions. Here, $26.2\%$ of examples were uncategorized by BERTopic while $2.5\%$ were uncategorized by LLooM.

\begin{figure*}[!tb]
  \includegraphics[width=1.0\textwidth]{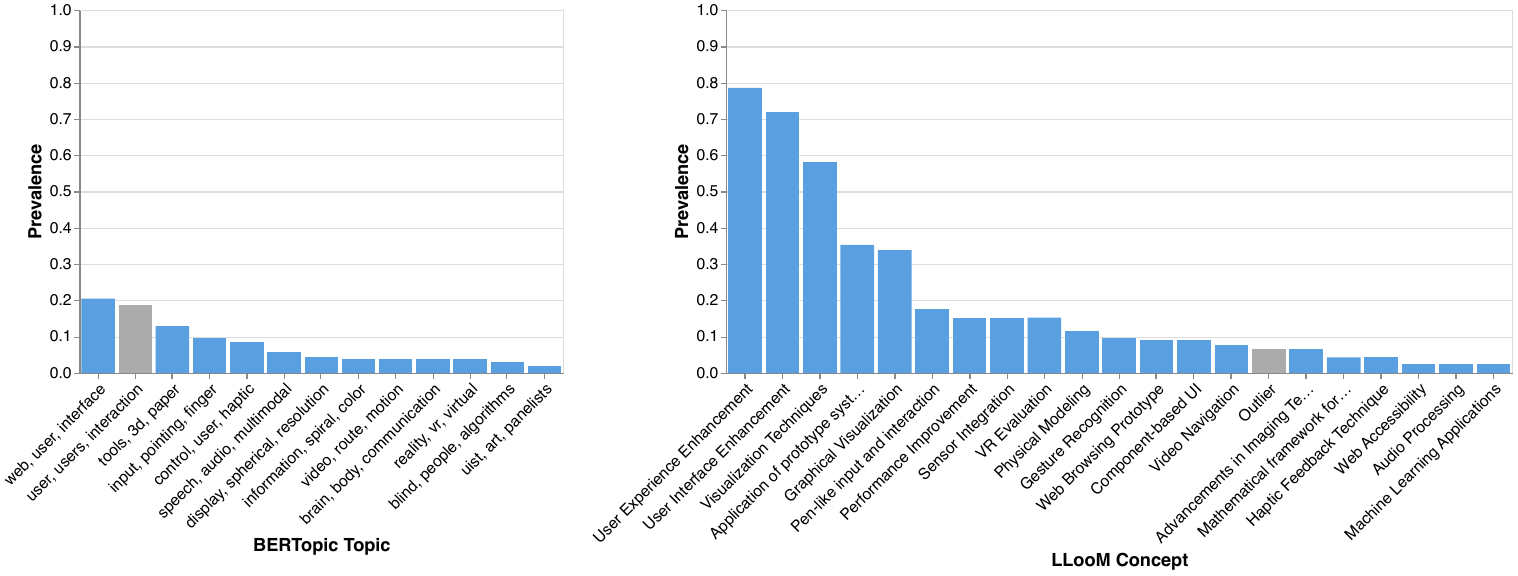}
  \caption{
   Scenario 3: HCI UIST papers dataset---BERTopic again places a sizeable portion of examples into an uncategorized set with $18.6\%$ of examples. LLooM concepts display a long tail distribution with a few high-frequency user interface concepts and a longer set of more nuanced concepts; 6.7\% of examples were outliers.
  }
  \label{fig:hist_uist}
  \Description{(Same histograms format as Figure~\ref{fig:hist_tox_fem}) BERTopic has a large uncategorized cluster with 18.6\% of examples. LLooM has 5 concepts that have more than 20\% of examples, and its outlier concept has 6.7\% of the examples.}
\end{figure*}

\subsection{Scenario 3: Analyzing UIST Paper Abstracts}
\label{section:hci_papers}
A recent large-scale literature review investigated the impact of HCI research on industry by analyzing patent citations~\cite{cao2023breaking}. This prior work used LDA topics to describe trends among research that influenced patents. We explore whether LLooM could help to characterize research from the past 30 years at major HCI venues with the same dataset of HCI paper abstracts.
We filter to those from UIST ($n=1733$) because the \citet{cao2023breaking} paper identified that UIST papers had an extremely outsized proportion of patent citations, and we sought to better understand the nature of UIST research over time and potential factors underlying its high industry impact. To enable comparisons across time periods, we gathered a stratified random sample across each decade from 1989-1998, 1999-2008, and 2009-2018 with $70$ papers from each decade for a total sample of $n=210$ papers for this exploratory analysis.

\subsubsection{Results}
LLooM generated $16$ distinct concepts, such as ``Gesture Recognition,'' ``Visualization Techniques,'' and ``Sensor Integration,'' shown in Figure~\ref{fig:hist_uist}. 
Meanwhile, BERTopic generated $12$ distinct topics with keywords such as ``control, user, haptic,'' ``reality, vr, virtual,'' and ``speech, audio, multimodal.''  
For this dataset, BERTopic outputs were more coherent than for the other scenarios, perhaps in part because academic abstracts are written to clearly signal their subject matter. Additionally, for this kind of analysis, low-level keywords are more useful than is typical since many keywords are precise technical terms (e.g., ``VR,'' ``haptics,'' and ``multimodal UIs.'') that are generally used in a standard, narrow sense.
Meanwhile, the LLooM concepts aligned quite strongly with the BERTopic topics, but areas of non-overlap appeared to surface several unique concepts. While most outputs were aligned with recognizable research topics, the concepts of ``Performance improvement,'' ``Prototype Systems,'' and ``Mathematical Frameworks'' appeared to characterize aspects of the work like the higher-level methods and evaluation strategies and all raised interesting questions about the common evaluation metrics and implementation approaches used at UIST compared to other HCI venues. By contrast, the non-overlapping BERTopic topics appeared to be additional research topic areas, but not new kinds of topics.
While $83.3\%$ of BERTopic results were covered by LLooM, $62.5\%$ of LLooM results were covered by BERTopic, so LLooM achieved somewhat higher coverage. Here, $18.6\%$ of examples were uncategorized by BERTopic while $6.7\%$ were uncategorized by LLooM.

\subsection{Scenario Limitations}
We note several limitations of these analysis scenarios. 
First, to provide a fairer comparison between LLooM and BERTopic, we only conducted one iteration of the LLooM algorithm. Then, because we prioritized interactive completion times for our scenarios, we sampled approximately 200 examples to use within LLooM for each scenario, but some of the datasets were much larger. Thus, there are risks that LLooM was not fully representative of the data and that its concepts could differ if run on a significantly larger dataset. However, we note that a benefit of LLooM's generated concept criteria is that even if concepts are induced from a smaller data sample, they can be applied to a much larger set to assess concept generalizability and coverage. 

We do not have manual annotations for the scenario datasets on ``ground truth'' concepts, so we cannot report on global coverage of LLooM concepts nor their alignment with manual analysts' generated concepts. We perform a ground truth concept coverage analysis in the next section, \S\ref{section:tech_eval}, with annotated datasets. Finally, while the scenarios were selected to span a variety of topic areas, dataset sizes, and analysis goals, LLooM results may differ when applied to other kinds of datasets.

\section{Technical Evaluations}
\label{section:tech_eval}

Next, we perform technical evaluations to compare LLooM concept generation against human annotations and state-of-the-art methods for unstructured text analysis. 
We investigate how well LLooM can generate concepts that recover ground truth concepts in two evaluations using (1)~real-world benchmark datasets drawn from Wikipedia articles and U.S. Congressional bills (\S\ref{section:concept_gen_benchmark_eval}) and (2)~a synthetic dataset for greater experimental control (\S\ref{section:concept_gen_synthetic}).
As in the LLooM scenarios, we include a BERTopic baseline as a state-of-the-art topic modeling method. Since this evaluation is performance-oriented, we add GPT-4 and GPT-4 Turbo baselines to understand how LLooM performs relative to base LLMs.

\subsection{Concept Generation: Benchmark Dataset}
\label{section:concept_gen_benchmark_eval}
First, we evaluate LLooM concept generation on real-world datasets drawn from prior work in topic modeling~\cite{pham2023topicgpt} that have unstructured text documents and human topic annotations: a Wikipedia articles dataset~\cite{merity2018regularizing} and a U.S. Congressional bills dataset~\cite{hoyle2022neuralTopicModels}. These annotations are explicitly defined as \textit{topics}, which tend to align with more generic concepts and may not fully capture the set of concepts that LLooM can generate. However, the topic annotations provide a helpful point of comparison with existing topic modeling methods.

\subsubsection{Metric}
The goal of concept induction with LLooM is to reliably surface informative, valid concepts from unstructured text. Thus, we assess the validity and comprehensiveness of LLooM's concepts by measuring how well they recover ground truth topics, which are generated by human annotators and known to occur in a given dataset. We use a metric of \textit{concept coverage} to assess how well LLooM and baseline methods recover ground truth concepts from a human-annotated dataset, whether that be a benchmark dataset or the synthetic dataset we describe in \S\ref{section:concept_gen_synthetic}.

For each method and dataset, we run 10 independent trials of concept generation for a total of 80 trials. Each trial randomly shuffles the dataset documents, uses new sessions for calls to the OpenAI API for LLooM and the GPT-4 variants, and trains a new topic model for BERTopic. 
For every trial, we determine \textit{coverage}, the proportion of ground truth topics that are covered by the generated concepts. We calculate automated coverage metrics using GPT-3.5 (\texttt{gpt-3.5-turbo}). Our few-shot prompt provides the ground truth and generated concepts and asks model to match each ground truth concept with at most one generated concept if its meaning matches the ground truth concept (Appendix~\ref{appendix:prompt_auto_coverage}).
To verify this automated coverage metric, we randomly sample the results of 16 trials (4 from each concept generation method) and manually match all ground truth and generated concepts for each trial. Treating the manual coverage as ground truth, we observe a mean absolute error (MAE) of 0.07 (i.e., an average case may have a manual coverage of 40\% and an automated coverage of 33\%).

\subsubsection{Method}
We evaluated four concept generation methods: LLooM, BERTopic, GPT-4, and GPT-4 Turbo. We use the same LLooM process and BERTopic setup described in \S\ref{section:lloom_scenarios}, but for parity with our GPT-4 baselines, we use GPT-4 for the \texttt{Synthesize} operator; we continue to use GPT-3.5 for the \texttt{Distill} operator steps. Additionally, we increase the input and output batch sizes of the LLooM \texttt{Cluster} and \texttt{Synthesize} operators to accommodate the longer documents of our benchmark datasets. We add baselines that directly query GPT-4 and GPT-4 Turbo with zero-shot prompts. For these baselines, we use the same prompt that underlies the LLooM \texttt{Synthesize} operator, but instead provide the full document text instead of the distilled and clustered text excerpts. Since GPT-4 has a limited context window, we randomly sample documents to fill the context window; all documents fit into the larger GPT-4 Turbo context window.

\subsubsection{Datasets}
The Wikipedia articles dataset (\texttt{Wiki}) consists of 14,290 articles and human annotations for 15 Generic topics, such as ``Art and architecture'' and ``Language and literature''. The Congressional Bills dataset (\texttt{Bills}) consists of 32,661 bill summaries and human annotations for 28 Generic topics, such as ``Education,'' ``Environment,'' and ``Health''. We use random samples of dataset documents (n=205 and n=213, respectively) stratified across topics, to accommodate context window limits for the GPT-4 baseline.
A downside of using publicly-available annotated datasets is that they may have appeared in the GPT pre-training corpus, which in part motivates our synthetic dataset evaluation. As prior work has noted, text-to-label mappings for the Wiki dataset may have appeared in the pre-training data~\cite{pham2023topicgpt}, so this dataset may present inflated estimates for the GPT-4 baselines. Meanwhile, the Bills dataset may provide a more realistic performance estimate: the data is less likely to have appeared in the GPT-4 training data since the bill summary texts and labels are stored separately.
The LLooM algorithm substantially transforms text spans before performing concept generation, so it likely does not ``benefit'' as greatly from GPT-4's potential knowledge of the Wiki dataset.

\begin{figure}[!tb]
  \includegraphics[width=0.65\linewidth]{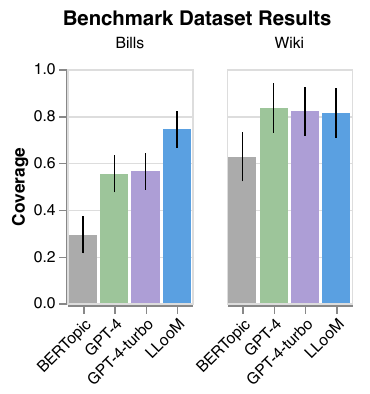}
  \caption{
    On the real-world benchmark datasets, LLooM exceeds baseline performance for the likely-unseen Bills dataset and matches GPT-4 baseline performance for the possibly-seen Wiki dataset, achieving coverage rates of $0.74$ and $0.81$ on the respective datasets.
  }
  \label{fig:concept_gen_eval_bench}
  \Description{The figure displays two histograms, with the Bills dataset on the left and the Wiki dataset on the right. The x-axis lists different methods (BERTopic, GPT-4, GPT-4 Turbo, and LLooM), and the y-axis has a 0-1 range and displays ground truth concept coverage. For the Bills dataset, BERTopic has the lowest coverage, followed by GPT-4 and GPT-4 Turbo with similar coverage, and LLooM has the highest coverage. For the Wiki dataset, BERTopic has the lowest coverage, and GPT-4, GPT-4 Turbo, and LLooM have similar higher coverage.}
\end{figure}

\subsubsection{Results}
LLooM exceeds baseline coverage by 17.9\% on the Bills dataset (LLooM: $M=0.74$, GPT-4 Turbo: $M=0.56$) and matches GPT-4 baselines on the Wiki dataset (LLooM: $M=0.81$, GPT-4: $M=0.83$, GPT-4 Turbo: $M=0.82$), as shown in Figure \ref{fig:concept_gen_eval_bench}. Supporting our note on the Wiki dataset's possible inclusion in the GPT pre-training data, GPT-4 and GPT-4 Turbo display substantially higher coverage on the Wiki dataset than the Bills dataset; the Wiki performance metrics may be inflated due to memorization of text-to-label mappings. Thus, it is promising that on the Bills dataset, LLooM maintains relatively consistent high coverage (only dropping 8.7\%), while GPT-4 Turbo coverage drops 25.6\%. In line with our LLooM scenarios, BERTopic displays substantially lower concept coverage for both datasets (Bills: $M=$0.29, Wiki: $M=$0.63) compared to the GPT-4 baselines and LLooM.

We further investigate these findings using a linear model with a fixed effect of method: \texttt{coverage \textasciitilde\ 1 + method}. We use a separate model for each dataset.
For the Bills dataset, we observe a significant main effect of method ($F(3, 36) = 22.36, p < .001$). A posthoc pairwise Tukey test finds statistically significant differences in coverage between all pairs of methods except for GPT-4 vs. GPT-4 Turbo ($p = 0.997$ for GPT-4 vs. GPT-4 Turbo, $p < .02$ for GPT-4 Turbo vs. LLooM, $p < .01$ for all other pairs). 
For the Wiki dataset, we also observe a significant main effect of method ($F(3, 36) = 3.568, p < .05$). A posthoc pairwise Tukey test only finds a statistically significant ($p < .05$) difference in coverage between BERTopic and GPT-4; there was no significant difference between any other pairs of methods.

We qualitatively compared the generated topics by inspecting all outputs for each method that matched a given ground truth topic (Tables~\ref{fig:tech_eval_outputs_wiki} and~\ref{fig:tech_eval_outputs_bills}). 
BERTopic topics were generally more vague (e.g., ``album, band, music'' for a ground truth Wiki \textit{music} topic or ``game, series, fantasy'' for a Wiki \textit{video games} topic). 
GPT-4 and GPT-4 Turbo topics often closely matched ground truth topics (e.g., ``Video Games'' for a Wiki \textit{video games} topic and ``Transportation Policy'' for a Bills \textit{transportation} topic), but GPT-4 displayed failure modes of combining multiple ground truth topics in a single topic (e.g., ``Artistic Works,'' which had a definition that mapped to Wiki \textit{music} or \textit{art and architecture} topics) while GPT-4 Turbo did not display this failure mode.
LLooM produced topics that matched closely with ground truth topics (e.g., ``Educational Policies'' for a Bills \textit{education} topic), but it also generated topics that highlighted \textit{other notable aspects} of content within a topic area (e.g., ``Community Development: Does the text discuss promoting education for community development?'' for the same Bills \textit{education} topic). For example, in a ground truth Wiki \textit{video games} topic, LLooM generated concepts like ``Video Game Discussion,'' ``Game Setting,'' and ``Character Design,'' and in a Wiki \textit{music} topic, LLooM generated concepts like ``Band Formation'' and ``Musician's Career.''

Overall, LLooM maintains high concept coverage on both datasets and provides substantial coverage benefits over baselines on the Bills dataset ($p < 0.02$).
GPT-4 Turbo is the nearest competitor on coverage metrics, but LLooM provides the added benefit of concepts that extend beyond matching ground truth labels to describe unique characteristics of data within a ground truth topic.

\begin{figure*}[!tb]
  \includegraphics[width=0.95\textwidth]{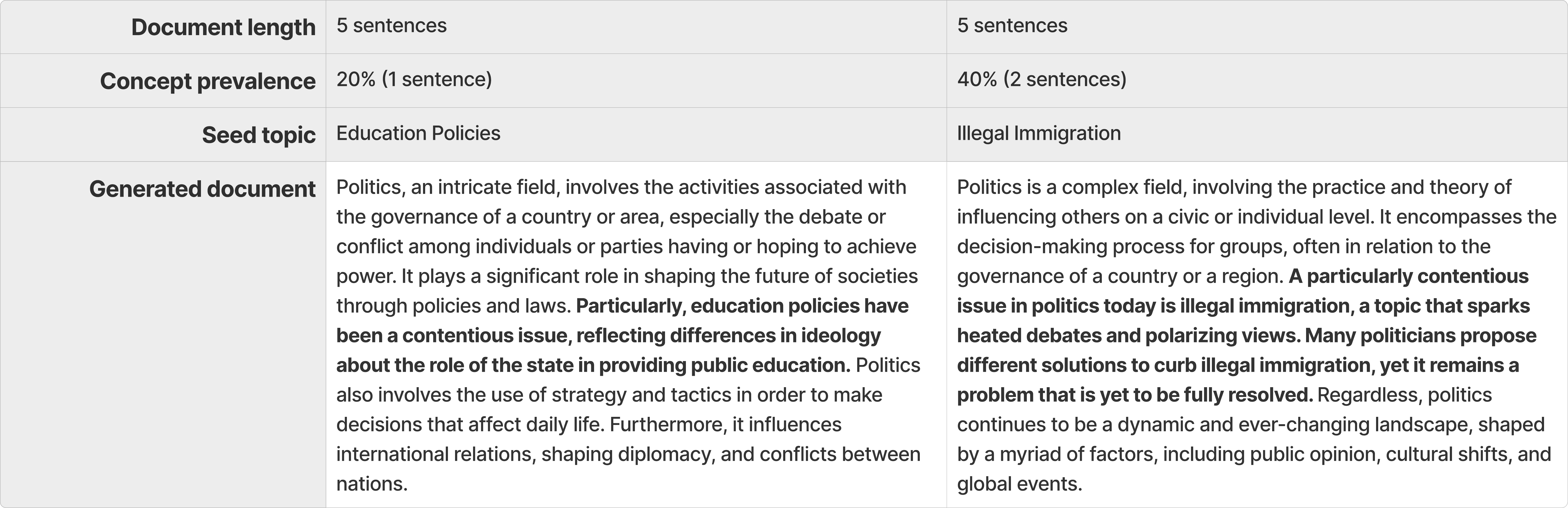}
  \caption{
    \textbf{Sample synthetic dataset documents}.
    We generated documents for combinations of document length, concept prevalence, and seed topic. The \textbf{bolded} portion indicates the seed concept sentences.
  }
  \label{fig:sample_gen_data}
  \Description{The figure displays a table with two columns. The first column has document length of 5 sentences, concept prevalence of 20\% (1 sentence), a seed topic of Education Policies, and a generated document. The second column has document length of 5 sentences, concept prevalence of 40\% (2 sentences), a seed topic of Illegal Immigration, and a generated document.}
\end{figure*}

\begin{figure}[!tb]
  \includegraphics[width=0.65\linewidth]{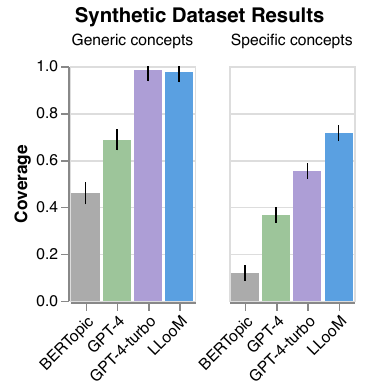}
  \caption{
    On the synthetic datasets, LLooM exceeds baselines on Specific concept coverage ($0.71$) and exceeds or matches baselines on Generic concept coverage ($0.98$).
  }
  \label{fig:concept_gen_eval_synth}
  \Description{(Same histogram format as Figure~\ref{fig:concept_gen_eval_bench}) There are two histograms: one for Generic concepts and one for Specific concepts. For Generic concepts, BERTopic again has the lowest concept coverage, followed by GPT-4, and GPT-4 Turbo and LLooM have similar high coverage. For Specific concepts, BERTopic has the lowest concept coverage, followed by GPT-4, followed by GPT-4 Turbo, and then LLooM has the highest coverage.}
\end{figure}

\subsection{Concept Generation: Synthetic Dataset}
\label{section:concept_gen_synthetic}
After demonstrating LLooM's performance on real-world datasets, we further probe its performance in a controlled setting.
Our synthetic dataset evaluation assesses how LLooM performs when we \textit{vary} the documents and concepts contained in a corpus.
Synthetic datasets grant us experimental control to independently study how performance is impacted by factors like document length and within-document concept prevalence, while holding constant the set of ground truth concepts and their across-document prevalence.  Additionally, since we construct these datasets, we can guarantee that these mappings of texts to ground truth labels do not occur in the GPT-4 pre-training data.

\subsubsection{Dataset generation}
Our synthetic dataset is generated from a seed set of ground truth Generic and Specific concepts that are held consistent, while we vary document length and within-document concept prevalence.

\textit{Parameters}.
First, we vary \textit{document length} since unstructured text can vary significantly in length depending on the domain (e.g., social media posts versus academic papers). Additionally, large language models like GPT-4 have limited context windows and display uneven performance across the context window~\cite{liu2023lost}. We test document lengths of 5 or 10 sentences; this approximately matches the range of document lengths in our LLooM scenarios (mean lengths of 2 to 8 sentences). 
Then, whether concepts comprise a small or large portion of a document, we still want LLooM to recover them since analysts are interested in both subtle and obvious concepts. Thus, we vary \textit{within-document concept prevalence}, operationalized as the percentage of sentences in the document related to a provided seed concept. We test concept prevalence values of 20\% or 40\%.
Finally, concepts are not monolithic: some concepts are lower-level, \textit{specific} ideas explicitly discussed in a document, while others are higher-level, more \textit{generic} themes that emerge from multiple lower-level concepts, and we want our method to capture both. While Generic concepts are useful in contexts like text clustering to surface overarching patterns, Specific concepts are useful in contexts like discourse analysis and can characterize nuanced patterns that inform theory-driven analysis. 
Thus, our dataset instantiates \textit{both Generic and Specific ground truth concepts}. 

\textit{Generation procedure}.
For our synthetic dataset, we chose an overall ``politics'' topic to align with politics-related datasets from our benchmark dataset evaluation (Bills dataset) and analysis scenarios (Partisan Animosity dataset).
We manually created a hierarchy of ten Generic concepts (e.g., ``Healthcare''), each of which has four constituent Specific concepts (e.g., ``Mental health,'' ``Health insurance''), all listed in Appendix~\ref{appendix:synth_data_concepts}.

For each unique combination of document length and concept prevalence, we generated 40 documents using GPT-4. Each document was generated by selecting one of the 40 Specific concepts, prompting the model to generate a document of \texttt{doc\_length} sentences about the overall ``politics'' topic, and requesting a fixed number of sentences related to the selected Specific concept based on \texttt{concept\_prevalence} (see sample generations in Figure~\ref{fig:sample_gen_data}).

\begin{lstlisting}[language=Markdown]
    Write a {doc_length}-sentence paragraph about 
    'politics'.
    In {concept_prevalence * doc_length} sentences of the paragraph, include content related to a SEED TOPIC '{low_level_concept}'. 
    Please only return a JSON with this format:
    {{
        "paragraph": "<PARAGRAPH>",
        "seed_topic_sentences": "<The sentences from PARAGRAPH related to SEED TOPIC>"
    }}
\end{lstlisting}

This approach allowed us to explicitly include Specific concepts in the text while implicitly invoking Generic concepts as themes that unify multiple Specific concepts. 

\textit{Verification}.
During the generation process, we programmatically verified that the total number of sentences in the documents matched the requested length and that the number of seed concept sentences aligned with the requested concept prevalence. We reviewed all documents and manually verified that the seed concept sentences sufficiently conveyed the specified concept. 

\subsubsection{Method}
We experimented with the same four methods---LLooM, BERTopic, GPT-4, and GPT-4 Turbo---using the same procedure as the benchmark dataset evaluation (Section \ref{section:concept_gen_benchmark_eval}).
For each combination of document length and concept prevalence, we evaluated each method on the corresponding set of synthetic documents with $n=10$ independent trials. We again calculated automated coverage metrics using GPT-3.5. We computed coverage for both Generic and Specific ground truth concepts.

\begin{figure*}[!tb]
  \includegraphics[width=0.65\textwidth]{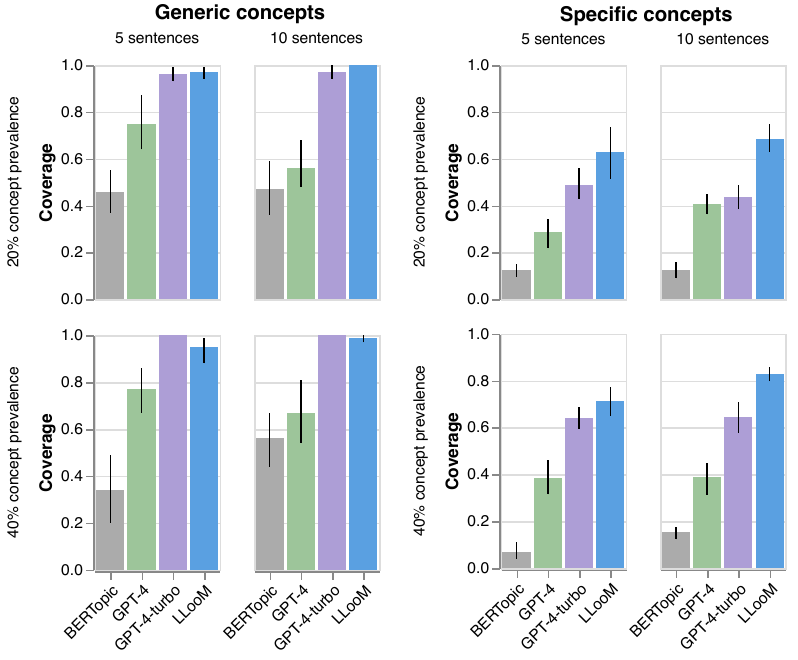}
  \caption{
    Synthetic dataset results by document parameters. Across document lengths and concept prevalence levels, LLooM achieves substantially higher Specific concept coverage and matches or exceeds Generic concept coverage compared to baselines.
  }
  \label{fig:synth_eval}
  \Description{ (Same histogram format as Figure 11, but with four of these plots) There are two rows and two columns of these histograms. The first row has plots for 20\% concept prevalence, the second row for 40\% concept prevalence; the left column has plots for Generic concepts and the right column has plots for Specific concepts. Each individual plot has two histograms, one for document lengths of 5 sentences and one for 10 sentences. Overall, we see similar trends among all of the histograms on the Generic concept side and the Specific concept side. For Generic concepts, BERTopic has the lowest coverage, followed by GPT-4, and then GPT-4 Turbo and LLooM have similar high coverage. For Specific concepts, BERTopic has the lowest concept coverage, followed by GPT-4, followed by GPT-4 Turbo, and then LLooM has the highest coverage.}
\end{figure*}

\subsubsection{Results}
Overall, we observe that LLooM achieves 16.0\% higher coverage than the nearest baselines on Specific concepts (LLooM: $M=0.71$, GPT-4 Turbo: $M=0.55$) and matches or exceeds baselines on Generic concepts (LLooM: $M=0.98$, GPT-4 Turbo: $M=0.98$, GPT-4: $M=0.69$, BERTopic: $M=0.46$), as shown in Figure~\ref{fig:concept_gen_eval_synth}. These trends are stable across document lengths and concept prevalence levels (Figure~\ref{fig:synth_eval}) and are consistent with our benchmark dataset findings, which have ground truth topics similar in form to Generic concepts. Notably, LLooM especially appears to provide benefit for Specific concepts and maintains high coverage while baseline methods substantially decline in coverage.

We analyze these results using a linear model with fixed effects of method, document length, and concept prevalence: \texttt{coverage \textasciitilde\ 1 + method + doc\_length + concept\_prevalence}. We use separate models for Generic concept coverage and Specific concept coverage.
For Specific concepts, we observe a significant main effect of method ($F(3, 154) = 227.4, p < .0001$), concept prevalence ($F(1, 154) = 22.0, p < .0001$), and document length ($F(1, 154) = 5.8, p < .05$). A posthoc pairwise Tukey test finds statistically significant differences in coverage between all pairs of methods ($p < .0001$), statistically significant differences between concept prevalence levels ($p < 0.0001$), and statistically significant differences between document lengths ($p < 0.05$). In other words, Specific concept coverage is highest for LLooM, then GPT-4 Turbo, then GPT-4, then BERTopic, and Specific concept coverage is higher for longer documents and those with higher concept prevalence.
For Generic concepts, we observe a significant main effect of method ($F(3, 154) = 115.03, p < .0001$). A posthoc pairwise Tukey test finds a statistically significant ($p < .0001$) difference in coverage between all pairs of methods except for GPT-4 Turbo vs. LLooM. Generic concept coverage is significantly higher for LLooM compared to GPT-4 and BERTopic, but not significantly different from GPT-4 Turbo.

We again compare the concepts generated by each method that successfully matched ground truth concepts (Table~\ref{fig:tech_eval_outputs_synth}). 
Again, BERTopic produces the most vague outputs (e.g., ``fiscal, economic, hoping'' for an \textit{economy} concept) that are supersets of Specific concepts.
Consistent with the benchmark datasets, GPT-4 and GPT-4 Turbo produce concepts that tend to align closely with Generic concepts (e.g. ``Healthcare Policy'' for a \textit{healthcare} concept). GPT-4 again displays an occasional failure mode of combining multiple ground truth concepts (e.g., ``Political Influence,'' which was defined in such a way that could map to \textit{economy} or \textit{foreign policy}), but GPT-4 Turbo does not appear to face this issue. Meanwhile, LLooM produces concepts that match both Generic as well as Specific ground truth concepts, as we saw for the benchmark dataset. For example, LLooM produces ``Economic Policies'' for an \textit{economy} concept, but it also produces concepts like ``Fiscal Measures'' and ``Economic Stability'' that are more specific and nuanced portrayals of data within the \textit{economy} concept.

In summary, LLooM performs strongly across all datasets, and it particularly excels relative to baseline methods for Specific concepts ($p < .0001$), where baseline performance suffers. LLooM, GPT-4, and GPT-4 Turbo produce competent Generic concepts, but LLooM is additionally able to recover Specific concepts in the dataset.

\subsection{Concept Classification}
We then evaluate LLooM's \texttt{Score} operator against human annotators (Appendix~\ref{section:concept_classif_eval}). LLooM attains inter-rater reliability ($\kappa = 0.63$, $\kappa = 0.645$) very similar to that of human annotators ($\kappa = 0.64$) and achieves moderate to high performance levels (Accuracy: $0.91$, Precision: $0.70$) on subjective concepts generated from our LLooM scenario datasets.

\section{Expert Case Studies} 
\label{section:eval}

Building on our analysis scenarios that showcase LLooM's concepts and our technical evaluation that supports the validity and coverage of these concepts, we explore how LLooM might aid \textit{realistic data analysis tasks} that go beyond the standalone task of concept generation.
We carry out first-use sessions with expert data analysts who have authored publications on two of our scenario datasets: (1)~Mitigating Partisan Animosity on Social Media and (2)~Analyzing the Industry Impact of HCI.
These sessions are intended as \textit{exploratory} probes to demonstrate how data analysts interact with LLooM concepts to make sense of their own data.
While the goal of the LLooM scenarios and technical evaluation was to validate LLooM outputs, the goal of the expert case studies was to surface design opportunities for the LLooM \textit{analysis experience} by highlighting preliminary differences from status quo data analysis tools. 
We focused on a small number of experienced analysts because they are a discerning and critical audience who may already hold strong understanding of a dataset, so they can provide expert feedback on the utility of LLooM outputs for data analysis.

\begin{figure*}[!tb]
  \includegraphics[width=0.6\textwidth]{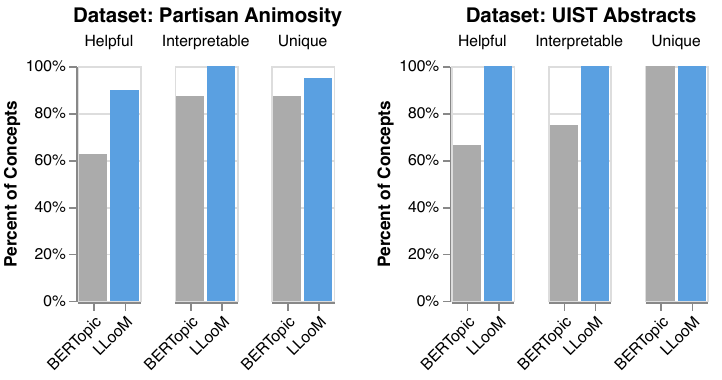}
  \caption{
    \textbf{Expert Analyst Assessments of Concept Quality}.
    Experts familiar with these datasets consistently rated a higher proportion of LLooM concepts as helpful, interpretable, and unique (non-overlapping).
  }
  \label{fig:expert_eval}
  \Description{The figure displays two sets of histograms, one on the left for the Partisan Animosity dataset and one on the right for the UIST Abstracts dataset. Within each side, there are three bar charts: one each for Helpful, Interpretable, and Unique labels. The x-axis has methods (BERTopic or LLooM), and the y-axis has the percentage of concepts that received the label. Across both datasets and all labels, BERTopic has a lower percentage than LLooM.}
\end{figure*}

Details on participant recruitment and session format are included in Appendix~\ref{appendix:expert_sess_study_design}. As a brief summary, each study consisted of a 1-hour session that included a BERTopic analysis task, a LLooM Workbench analysis task, and a concluding interview. During the session, participants engaged in a think-aloud protocol as they conducted exploratory data analysis of the same dataset that they had analyzed for a prior publication.

\subsection{Expert 1: Mitigating Partisan Animosity on Social Media}
In the first session, the LLooM Workbench helped the expert analyst to identify previously-unnoticed trends and activated relevant domain knowledge to inspire theory-driven analyses.
For the BERTopic topics, the analyst labeled 5 as helpful ($62.5\%$), one as uninterpretable ($12.5\%$), and one as overlapping with another topic ($12.5\%$), as shown in Figure~\ref{fig:expert_eval}. For LLooM concepts, the analyst labeled 18 as helpful ($90\%$), none as uninterpretable ($0\%$), and one as overlapping with another concept ($5\%$).

\subsubsection{BERTopic Analysis Process---Making sense of vague and overlapping topics}
The analyst reviewed topic keywords (e.g., ``oil, gas, energy, strategic'') and attempted to explain each topic (e.g., \textit{Natural resources and energy}) based on prior knowledge of the dataset. They spent time exploring examples primarily to compare two highly similar topics (``house, republicans, democrats'' and ``rep, congressman, great''), but could not identify a meaningful difference.

\subsubsection{LLooM Analysis process---Exploring data through the lens of concepts}
By contrast, with the LLooM Workbench, the analyst did not need to spend time interpreting each concept and primarily spent time inspecting the data \textit{through the lens} of the concept. 

\textit{Exploring concepts that match or violate expectations}. The analyst selectively explored concepts that \textit{differentiated} low and high partisan animosity examples based on the concept prevalence histograms. 
Several concepts matched the analyst's expectations as associated with high partisan animosity (e.g., ``Government-Related Themes'' and ``Political Commentary'') or low partisan animosity (e.g., ``Government Accountability'' and ``Public Health Concern''). However, LLooM helped the analyst to discover an \textit{unanticipated} and particularly helpful ``Political Party Positions'' concept that was prevalent among high partisan animosity posts and surfaced a pattern of attacks on out-party stances.

\textit{Investigating nascent patterns}. 
Starting from an \textit{existing} ``Policy-related'' concept, the analyst noticed a pattern of posts dramatizing the impact of particular policies (e.g., immigration and border policies). They explored this pattern further by creating a \textit{variant of the original concept} named ``Crisis'' with the criteria, ``Does this example mention crisis due to a policy?'' In a few seconds, they were pleased to see that they had successfully identified a salient cluster of posts that carried high partisan animosity scores.

\textit{Activating relevant domain knowledge}. 
Prompted by this exploration, the analyst was reminded of their domain knowledge on anti-democratic attitudes in political science literature \cite{voelkel2023megastudy}, which included \textit{social distrust}. They created a new concept named ``Social Distrust'' with the criteria, ``Does this example display distrust of other people or society?''
The analyst found that these examples received mid-to-high partisan animosity scores, but did not fall in the highest bucket of scores, so perhaps that factor was less predictive of the most severe cases of partisan animosity.
While it would ordinarily be challenging to extract examples that display social distrust, which manifests implicitly rather than explicitly, LLooM allowed the analyst to successfully capture the concept.

\subsubsection{Interview Takeaways}
Overall, while BERTopic allowed the analyst to see data in terms of loose groupings, LLooM allowed them to \textit{navigate} and \textit{understand} data in terms of meaningful concepts.

\textit{BERTopic is a map, LLooM is a vehicle}. 
BERTopic topics helped the analyst to ``visualize the main patterns.''
They felt that for future qualitative coding, topics like these could simplify their work because examples within each cluster would likely have similar ratings for constructs like partisan animosity.
With LLooM Workbench, the analyst felt that the system ``[did] a much better job in terms of visualizing and helping me navigate concepts as well as examples under those concepts.'' 

\textit{LLooM may aid preliminary phases of qualitative analysis}. 
The analyst expressed that the LLooM Workbench would ``help [them] a lot in providing guidance on different categorizations of the data'' for qualitative analysis. 
They raised a potential concern that LLooM's outputs could impact their judgment in categorizing data: since it ``already gives me an initial categorization, it might affect my judgement.'' 
However, ``given how precise the concepts are,'' they felt that as a first step of coding, LLooM would be extremely helpful to save time and grant a better understanding of the whole dataset, especially for large datasets.

\subsection{Expert 2: Analyzing UIST Paper Abstracts}
LLooM Workbench helped the second analyst to actively explore hypotheses and carry out analysis ideas that were previously challenging to enact.
For the BERTopic topics, they labeled 8 as helpful ($66.7\%$), 3 as uninterpretable ($25\%$), and none as overlapping with another topic ($0\%$), as shown in Figure~\ref{fig:expert_eval}. For LLooM concepts, the analyst labeled all 16 as helpful ($100\%$), none as uninterpretable ($0\%$), and none that were overlapping with each other ($0\%$).

\subsubsection{BERTopic Analysis Process---Dealing with incoherent and overly-generic topics}
The second analyst spent most of their time reviewing the BERTopic keywords and only inspected examples to make sense of topics with uninterpretable keywords. 
They primarily looked for coherent groups of terms within the keyword sets, such as ``reality, vr, virtual,'' but struggled to author manual labels for 3 of the topics ($25\%$). 

\textit{Difficulties iterating on uninformative topics}. Several clusters consisted of terms like ``user'' and ``interface'' that might be informative in a general sense, but were uninformative in this analysis context. Given the ubiquity of users and interaction in HCI research, such clusters didn't help the analyst to understand the patterns happening \textit{within} a conference like UIST.
This was a major painpoint when they had previously used LDA for topic modeling on this dataset, as they had to perform multiple rounds of iteration to catch stopwords and optimize output clusters, which was time-consuming and caused them to doubt whether their results were robust.

\subsubsection{LLooM Analysis Process---Leveraging concepts to explore hypotheses}
When using the LLooM Workbench, the analyst noted that it contrasted sharply with their prior experience with traditional topic models. 

\textit{Less time validating, more time exploring}. 
With LLooM, they were able to immediately understand the extracted concepts and verify how they mapped to specific documents. The analyst deemed all of the LLooM concepts as both interpretable and helpful for their analysis task of understanding research at UIST, and they found the criteria prompt especially helpful in clarifying the meaning of concepts.
Most of the analyst's time was spent \textit{using} the concepts to compare changes in paper topics or methods over the decades.

\textit{Exploring their own hunches and analysis ideas}. 
The analyst was particularly excited about authoring new concepts with LLooM, as this was a barrier with traditional topic modeling tools where analysts cannot proactively specify their \textit{own} topics that they wish to explore. The analyst was curious about whether more HCI researchers were incorporating AI into their systems, since this appeared to be the case from their anecdotal experience. They authored a new concept called ``AI'' with the criteria ``Does this example include concepts of artificial intelligence?'' and indeed found that there was a steady rise in AI-related papers across the decades. 

\textit{Investigating concepts that are challenging to describe}. 
In past analyses where the analyst had a hypothesis and wanted to ``zoom in'' on that phenomenon, they had to rely on keyword search, which was time-intensive, required domain knowledge, and could result in coverage gaps. They felt that LLooM would be highly useful for these analysis tasks not only to lower effort, but to increase coverage. LLooM successfully surfaced examples in the AI concept that didn't explicitly use the AI term, such as a paper that only mentioned ``object recognition,'' and the analyst commented that even researchers in the field would likely struggle to come up with terms like this before diving into the data.

\subsubsection{Interview Takeaways}
In summary, the analyst found LLooM helpful in not only providing a ``straightforward, high-level idea'' of data, but also fostering \textit{proactive} analyst-led data explorations.

\textit{LLooM should help analysts calibrate their trust}. One limitation that they raised was that data scientists and computational social scientists would likely want to have quantitative metrics to indicate the robustness and reliability of the tool to increase their confidence in building on the output concepts. Additionally, users in these domains would likely want to better understand LLooM's internal process to calibrate their trust in the tool. 

\textit{LLooM can facilitate theory-driven analysis}. The analyst was most enthusiastic about the possibility for the tool to support more \textit{theory-driven} analyses in response to LLooM's automatically extracted concepts. While they had wanted to analyze data in this way in prior research projects, it was challenging to execute this kind of analysis with existing tools.

\section{Discussion}
In this paper, we present LLooM, a concept induction algorithm that extracts high-level, interpretable concepts from unstructured text datasets. LLooM not only improves topic quality and coverage, but also provides benefits to steerability and interpretability. Here, we discuss design implications, limitations, and opportunities for future work.

\subsection{Design Implications}
LLooM points toward several design opportunities in the realms of topic modeling and interactive data analysis.

\subsubsection{Redesigning data analysis abstractions to support theory-driven analysis}
With LLooM, we ask whether it is possible to redesign the core abstractions of our data analysis systems to center around the way analysts would like to think about their data. Based on our evaluations and preliminary findings, it appears that it is indeed possible to orient a topic modeling process entirely around human-understandable concepts expressed in natural language, and enable analysts to steer the model's attention toward specific analytic goals. By linking data-driven results with human-readable ideas, we can enact a very different data analysis experience where an analyst can ``read'' emergent patterns from data and, in response, ``write'' their theory to apply it back onto the data.
    
\subsubsection{Introducing automation to aid reflection on analysis processes}
By automating elements of the data analysis process, we can free analysts to step back one level and not just enact their analysis process, but reflect and identify potential gaps therein. Moreover, in contexts such as computational social science, analysts may need to make credible commitments for replicability and generalizability purposes that they have not overly biased the analysis process. 
In these cases, LLooM can automatically carry out key aspects of manual data analysis, such as distilling data, grouping together relevant items, synthesizing trends into concepts, and applying those concepts to categorize data. LLooM can aid reflection by guiding users to clarify the meaning of concepts, catch blindspots in their analysis that aren't covered by concepts, and initiate parallel re-runs to explore a variety of data interpretations.
In contrast, if the analyst \textit{does} wish to inject their insight and perspectives into the analysis, as is more common in ethnomethodological traditions, LLooM can operate in a closed loop with the analyst.
    
\subsubsection{Innovating on our core algorithmic operators}
To implement LLooM, we combined the core operators introduced in this work (e.g., Distill, Cluster, and Synthesize) into an architecture that drew inspiration from the qualitative analysis process. However, there is a much broader design space of operators and implementations. We see exciting opportunities to dynamically rearrange and restructure these operators as building blocks for different analysis tasks depending on an analyst's goals. Going further, we could innovate new operators that align with the cognitive processes of not just data analysts, but other human domain experts for tasks beyond data analysis.

\subsection{Limitations and Future Work}
LLooM also presents critical design challenges, especially given its use of large language model outputs and its specific use of OpenAI's GPT models. These point to important future work directions.

\subsubsection{Uncertain LLM behaviors: risks of uneven cross-domain performance}
One core limitation of this work, and any work that builds upon large language models, is that we currently lack reliability and performance guarantees. 
LLM performance can vary widely across domains and greatly depends on the training data, which is often withheld from public knowledge. While we can expect LLMs like GPT-4 to perform strongly on text similar to the distribution of large-scale Internet text data on which they were trained, performance may decline in specialized domains such as law, medicine, and fields requiring technical expertise.
Novel techniques may be needed to enable concept induction in areas underrepresented in LLM training data.
LLMs often err in following instructions, struggle with logical statements, or produce outputs with hallucinations that are not faithful to the original data. We cannot entirely remove the possibility of such foundational errors, but our system additionally mitigates the risk of downstream harm by heavily incorporating human review: analysts can trace concepts back to lower-level concepts and original data examples, and they can review concept scores and rationales to catch when models fail.

\subsubsection{Drawbacks of closed-source LLMs: cost and lack of transparency}
Compounded on the uncertainties of large language models in general, there are additional downsides of closed-source models like OpenAI's GPT models, which we use in our LLooM implementation. 
Since we lack transparency on both the data on which these models were trained and the design of the models themselves, we have limited ability to anticipate blindspots that would impact LLooM's functionality. 
Additionally, the use of OpenAI models presents barriers to reproducibility: the model versions underlying the APIs may change at any time without our knowledge, and we lack the control to invoke the same model version we may have used in the past. 
We opt to use the closed-source OpenAI GPT models because they represent the state-of-the-art; our preliminary testing with other models could not reliably execute the synthesis operations central to our approach. However, as open-source model capabilities improve, future work should explore strategies for using open-source models for concept induction.

Another limitation of closed-source LLMs is that it is costly to run our process at extremely large scales since our method depends on calls to external APIs that charge by token usage and that enforce token limits. 
In the years since the original releases of APIs for LLMs, costs have already dramatically decreased, so we anticipate that cost and efficiency issues will become less of a barrier in the future. Given that concept scoring is an especially costly part of the pipeline, if analysts need to scale up classification, they could explore training distilled models using a smaller set of LLM-labeled examples to reduce the cost and speed of inference, or drawing on open-source LLMs.

\subsubsection{Potential to bias analysts}
Lastly, as surfaced by our expert case studies and in prior literature on AI-assisted data analysis~\cite{jiang2021serendipity, hong2022scholastic}, AI-based analysis tools like LLooM may risk biasing analysts or limiting their agency to lead analyses. 
If analysts too heavily depend on LLooM outputs---by not inspecting the concepts, not exploring potential gaps outside of the set of generated concepts, or overrelying on the automated concept scores---they may miss important patterns in the data or may inadvertently build on low-quality or faulty model outputs.
Thus, future work should help users to calibrate their trust in LLooM with indicators of reliability and potential knowledge gaps. This work should further aid users in verifying system outputs, manually inspecting results, and leading follow-up analyses to augment exploratory LLooM analyses.
Along this line, an important limitation of LLM tools is that the values and biases encoded in LLMs are unclear, but they certainly can shape the concepts that our system generates. Future tools need to design around this challenge and provide greater transparency and control about the values embedded in LLM-led data analysis.

\section{Conclusion}
Unstructured text holds a vast amount of information, but it remains difficult to derive meaningful insights from data in this form. 
It is especially challenging to enact \textit{theory-driven analyses} of unstructured text.
Current tools like topic modeling and clustering are helpful, but tend to output surface features like ``rep, congressman, great'' that require substantial effort to interpret and validate.
We introduce the task of \textit{concept induction}, a computational process that takes in unstructured text and produces high-level concepts---human-interpretable descriptions defined by explicit \textit{inclusion criteria} (e.g., a ``Government and community collaboration'' concept defined by criteria like ``Does the text example mention a government program or initiative and community engagement or participation?'').
High-level concepts provide the affordances to ``read'' out data patterns in an interpretable form and to ``write'' out actionable theories that can be applied back to data.
We present LLooM, a concept induction algorithm that implements a novel LLM-powered \texttt{Synthesize} operator to iteratively sample unstructured text and propose high-level concepts of increasing generality.
By instantiating LLooM in a mixed-initiative text analysis tool called the LLooM Workbench, we demonstrate that its concepts are able to exceed the quality of topic models. With LLooM, analysts can see and interact with data in terms of interpretable, actionable concepts to lead theory-driven analyses of unstructured text.

%%
%% The acknowledgments section is defined using the "acks" environment
%% (and NOT an unnumbered section). This ensures the proper
%% identification of the section in the article metadata, and the
%% consistent spelling of the heading.
\begin{acks}
We thank our anonymous reviewers in addition to Omar Shaikh, Jordan Troutman, and Farnaz Jahanbakhsh for their valuable feedback on our paper. We thank Zachary Xi for contributions to our evaluations. This work was supported in part by IBM as a founding member of the Stanford Institute for Human-centered Artificial Intelligence (HAI) and by NSF award IIS-1901386. Michelle S. Lam was supported by a Stanford Interdisciplinary Graduate Fellowship.
\end{acks}

%TC:ignore

%%
%% The next two lines define the bibliography style to be used, and
%% the bibliography file.
\bibliographystyle{ACM-Reference-Format}
\bibliography{references}

%%
%% If your work has an appendix, this is the place to put it.
\appendix
\section{Prompts}
\label{appendix:prompts}

\subsection{\texttt{Distill} operator: Filter step prompt}
\label{prompt:distill_filter}
\begin{lstlisting}[language=Markdown]
    I have the following TEXT EXAMPLE:
    {text_example_json}
    
    Please extract {n_quotes} QUOTES exactly copied from this EXAMPLE {seed_phrase}. 
    Please respond ONLY with a valid JSON in the following format:
    {{
        "relevant_quotes": [ "<QUOTE_1>", "<QUOTE_2>", ... ]
    }}
\end{lstlisting}

\subsection{\texttt{Distill} operator: Summarize step prompt}
\label{prompt:distill_summarize}
\begin{lstlisting}[language=Markdown]
    I have the following TEXT EXAMPLE:
    {text_example_json}
    
    Please summarize the main point of this EXAMPLE {seed_phrase} into 
    {n_bullets} bullet points, where each bullet point is a {n_words} word phrase. 
    Please respond ONLY with a valid JSON in the following format:
    {{
        "bullets": [ "<BULLET_1>", "<BULLET_2>", ... ]
    }}
\end{lstlisting}

\subsection{\texttt{Synthesize} operator prompt}
\begin{lstlisting}[language=Markdown]
    I have this set of bullet point summaries of text examples:
    {bullets_json}
    
    Please write a summary of {n_concepts} unifying patterns for these examples {seed_phrase}.
    For each high-level pattern, write a {n_name_words} word NAME for the pattern 
    and an associated 1-sentence ChatGPT PROMPT that could take in a new text example 
    and determine whether the relevant pattern applies. 
    Please also include {n_example_ids} example_ids for items that BEST exemplify the pattern. 
    Please respond ONLY with a valid JSON in the following format:
    {{
        "patterns": [ 
            {{
                "name": "<PATTERN_NAME_1>", 
                "prompt": "<PATTERN_PROMPT_1>", 
                "example_ids": ["<EXAMPLE_ID_1>", "<EXAMPLE_ID_2>"]
            }},
            {{
                "name": "<PATTERN_NAME_2>", 
                "prompt": "<PATTERN_PROMPT_2>", 
                "example_ids": ["<EXAMPLE_ID_1>", "<EXAMPLE_ID_2>"]
            }},
        ]
    }}
\end{lstlisting}

\subsection{\texttt{Score} operator prompt}
\begin{lstlisting}[language=Markdown]
    CONTEXT: 
        I have the following text examples in a JSON:
        {examples_json}
    
        I also have a pattern named {concept_name} with the following PROMPT: 
        {concept_prompt}
    
    TASK:
        For each example, please evaluate the PROMPT by generating RATIONALE of your thought process
        and providing a resulting ANSWER of ONE of the following multiple-choice options, including just the letter: 
        - A: Strongly agree
        - B: Agree
        - C: Neither agree nor disagree
        - D: Disagree
        - E: Strongly disagree
        Respond with ONLY a JSON with the following format, escaping any quotes within strings with a backslash:
        {{
            "pattern_results": [
                {{
                    "example_id": "<example_id>",
                    "rationale": "<rationale>",
                    "answer": "<answer>",
                }}
            ]
        }}
\end{lstlisting}

\subsection{Automated coverage prompt}
\label{appendix:prompt_auto_coverage}
\begin{lstlisting}[language=Markdown]
    I have this set of CONCEPTS:
    {ground_truth_concepts}
    
    I have this set of TEXTS: 
    {generated_concepts}
    
    Please match at most ONE TEXT to each CONCEPT. To perform a match, the text must 
    EXACTLY match the meaning of the concept. 
    Do NOT match the same TEXT to multiple CONCEPTS.
    
    Here are examples of VALID matches:
    - Global Diplomacy, International Relations; 
    rationale: "The text is about diplomacy between countries."
    - Statistical Data, Quantitative Evidence; 
    rationale: "The text is about data and quantitative measures."
    - Policy and Regulation, Policy issues and legislation; 
    rationale: "The text is about policy, laws, and legislation."
    
    Here are examples of INVALID matches:
    - Reputation Impact, Immigration
    - Environment, Politics and Law
    - Interdisciplinary Politics, Economy
    
    If there are no valid matches, please EXCLUDE the concept from the list. 
    Please provide a 1-sentence RATIONALE for your decision for any matches. 
    Please respond with a list of each concept and either the item it matches or NONE 
    if no item matches in this format:
    {{
        "concept_matches": [
            {{
                "concept_id": "<concept_id_number>",
                "item_id": "<item_id_number or NONE>",
                "rationale": "<rationale for match>",
            }}
        ]
    }}
\end{lstlisting}

\begin{figure*}[!tb]
  \includegraphics[width=0.75\textwidth]{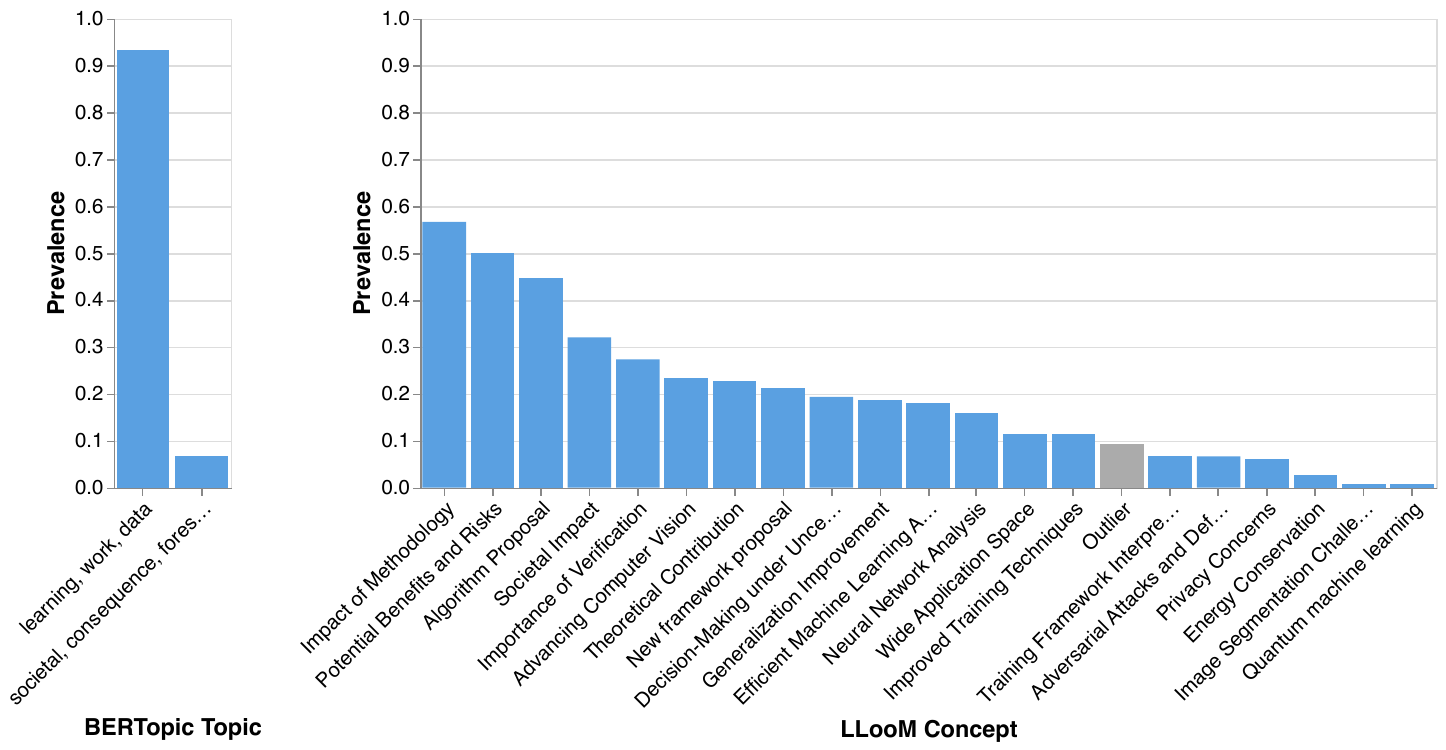}
  \caption{
    \textbf{NeurIPS Broader Impact Statements, Topic Prevalence}.
    BERTopic struggles with only two categories, one of which appears to be a vague catch-all topic with $93.3\%$ of examples. LLooM surfaces concepts that range from characterizing the majority of data to small subsets, and only fails to categorize $9.3\%$ of examples.
  }
  \label{fig:hist_neurips}
  \Description{(Same histograms format as Figure~\ref{fig:hist_tox_fem}) BERTopic has one large cluster with 93.3\% of examples and a smaller cluster with 0.7\% of examples. LLooM has 8 concepts that contain more than 20\% of examples, and it has an outlier concept that has 9.3\% of examples.}
\end{figure*}

\section{Additional Methods}
\label{appendix:addl_methods}

\subsection{Expert Case Study: Study Design}
\label{appendix:expert_sess_study_design}
The Expert Case Study required participants who have expertise in data analysis: specifically, those who have conducted an analysis of unstructured text documents. It was important that they had already conducted this analysis (so that they had enough prior knowledge of the data to distinguish helpful and unhelpful concepts) and that the dataset could be shared publicly (since the analysis scenarios and expert case studies would be published). Thus, our eligibility criteria were (1)~that the analyst had previously authored an academic publication based on a dataset and (2)~that the data consisted of unstructured text documents. For our exploratory analysis goals, we recruited $N=2$ participants through contacts in the university setting. Expert 1 was a postdoctoral scholar in Communication and Human-Computer Interaction with research interests in emerging media technologies and human-centered AI. Expert 2 was a Ph.D. student in Human-Computer Interaction and Natural Language Processing with research interests in computational social science and large-scale data mining. The participants had no knowledge of the LLooM Workbench and its functionality prior to the study session.

For the BERTopic analysis task, the participant was given a spreadsheet view populated with BERTopic outputs for their dataset. A summary tab displayed the keywords and size of each topic; a detail tab displayed a filterable view with all documents and their assigned topic. To understand how the expert interpreted the topics, we first had them complete a \textit{naming} task of providing a meaningful name for each topic. Then, the participant was asked to freely explore the data and topics. Finally, we had them complete an \textit{annotation} task on whether each topic was helpful (aids their understanding of the dataset), interpretable (has a discernible meaning), and unique (does not share the same meaning as another topic).
For the LLooM analysis task, the participant accessed the LLooM Workbench via a computational notebook already populated with the LLooM-generated concepts for their dataset. The participant was asked to review the generated concepts, and then to freely explore the data based on their interests. Towards the end of this section, we asked the participant to complete a \textit{concept modification} task to either edit or add one new concept. To conclude, we had them complete the same \textit{annotation} task on LLooM concepts.

The session was roughly split into 5 minutes for consent and setup, 15 minutes for analysis using BERTopic, 5 minutes for a post-interview on BERTopic, 5 minutes for a LLooM Workbench tutorial, 15 minutes for analysis using LLooM Workbench, and 10 minutes for a final interview on LLooM and their overall experience with both tools.
Each session was conducted remotely over a video call, and participants were compensated with a \$45 Amazon gift card. 

\section{Additional Results}
\label{appendix:addl_results}

\subsection{Scenario 4: Investigating Anticipated Consequences of AI Research}
\label{appendix:case_study_neurips}
In 2020, NeurIPS, a premier machine learning research conference, required authors to include a broader impact statement in their submission in an effort to encourage researchers to consider negative consequences of their work. These statements provide a  window into the ethical thought processes of a broad swath of AI researchers, and prior work has performed a qualitative thematic analysis on a sample of 300 statements ~\cite{nanayakkara2021impactStatements}. Using this dataset, we explore how LLooM might help us to understand how AI researchers discuss downstream consequences, ethical issues, and potential mitigations.

\subsubsection{Results}
LLooM generated $14$ unique concepts, including examples like ``Adversarial Attacks and Defenses,'' ``Privacy Concerns,'' and ``Energy Conservation,'' as shown in Figure \ref{fig:hist_neurips}. In contrast, BERTopic generated only $2$ topics with keywords such as ``societal, consequences, foreseeable'' and ``learning, work, data.'' 
The BERTopic topics were all quite generic (our manual analysis mapped the topics to labels of ``Machine Learning Techniques'' and ``Ethics and Societal Impacts''). Since these topics could likely apply as a category label for all impact statements, they do not help analysts to break down the data into emergent trends. The LLooM results also included some more generic concepts (e.g., ``'Societal Impact'), but it also identified specific \textit{kinds of impact} mentioned in statements, including both positive impacts  (e.g., ``Energy Conservation,'' ``Generalization Improvement,'' ``Improved Training Techniques,'' and ``Efficient ML Algorithms'') and negative impacts (e.g., ``Privacy Concerns,'' ``Adversarial Attacks''). Furthermore, the concepts encapsulated \textit{proposed solutions} to downstream impacts of AI research (e.g., ``Adversarial Defenses,'' ``Importance of Verification'').

While $100\%$ of BERTopic results overlapped with LLooM, only $14.3\%$ of LLooM results overlapped with BERTopic, so there was a substantial portion of LLooM concepts that were novel contributions. Here, none of examples were uncategorized by BERTopic while $9.3\%$ were uncategorized by LLooM. However, one of the two BERTopic results (``learning, work, data'') appears to be a vague catch-all topic; BERTopic assigned $93.3\%$ of examples to this group.

\subsection{Concept Classification Evaluation}
\label{section:concept_classif_eval}
We perform an additional evaluation on the reliability of LLooM's automated concept classification with the \texttt{Score} operator. To assess how well LLooM aligns with human judgment, we sample LLooM-generated concepts, gather human annotations on documents for each concept, and compare the results with LLooM scores.

\subsubsection{Method}
For this evaluation, we sample concepts from the four LLooM scenario datasets. To capture the system's performance on both rare and common concepts, we perform a stratified random sample based on concept prevalence, the proportion of documents that LLooM classified as matching a concept.\footnote{We only conservatively classify examples as positive only if they receive an annotation of ``strongly agree,'' the most confident label option. All other label options are considered negative.} 
For each dataset, we sampled one concept from each quartile of concept prevalence for a total of four concepts. Then, for each selected concept, we constructed balanced datasets with $n=100$ documents by taking a stratified random sample of 50 positive documents (those that were classified as matching the concept) and 50 negative documents. For rare concepts with fewer than 50 positive documents, the remainder was drawn from a random sample of negative documents.

Included below are the sampled concepts for each dataset:
\begin{itemize}
    \item Partisan Animosity dataset:
    \begin{itemize}
        \small
        \item Advocacy: Does the text example advocate for a cause or issue?
        \item Event: Is this text example related to an event?
        \item Political Party Positions: Does the text example mention the positions or actions of political parties?
        \item Social Justice Focus: Does the text example emphasize working towards a just future?
    \end{itemize}
    \item Toxic Content dataset:
    \begin{itemize}
        \small
        \item Expressing Frustration: Does the text example involve expressing frustration or disbelief?
        \item Men's Perception of Unfair Treatment: Does the text example discuss men feeling treated unfairly in society?
        \item Seeking Explanation: Does the text example seek an explanation for a certain behavior?
        \item Stereotyping Women: Does the text example involve stereotyping women?
    \end{itemize}
    \item UIST Abstracts dataset:
    \begin{itemize}
        \small
        \item Application of Prototype System: Does the text example discuss the application of a prototype system to various interfaces?
        \item Pen-like Input and Interaction: Does the text example involve precise pen-like input and handle interaction?
        \item User Experience Enhancement: Does the example describe a product or technology that enriches the user's experience?
        \item VR Evaluation: Does the text example involve evaluating and improving immersion in VR?
    \end{itemize}
    \item NeurIPS Statements dataset:
    \begin{itemize}
        \small
        \item Importance of Verification: Does the text example emphasize the importance of verifying data or systems?
        \item New Framework Proposal: Does the text example propose a new framework?
        \item Potential Benefits and Risks: Does the example discuss potential benefits and risks?
        \item Wide Application Space: Does the example mention wide application space for generic objects?
    \end{itemize}
\end{itemize}

To assess inter-rater reliability, two members of the research team independently annotated the four sampled concepts for one dataset (the Partisan Animosity dataset), each annotating 400 documents in total. One rater annotated the documents for the remaining three datasets. For each document, based on the concept name and inclusion criteria, each annotator selected from the same multiple-choice options provided to GPT-4 in the LLooM \texttt{Synthesize} operator prompt, ranging from whether they ``strongly agree'' to ``strongly disagree'' that the document matches the concept. 
Then, we compare these manual scores with those generated by LLooM in the concept scoring step. For inter-rater reliability, we use Cohen's $\kappa$ because we only consider pairs of raters, our scale is categorical (binary labels), and our data is approximately balanced.

\subsubsection{Results}
\begin{figure*}[!bt]
  \includegraphics[width=0.7\textwidth]{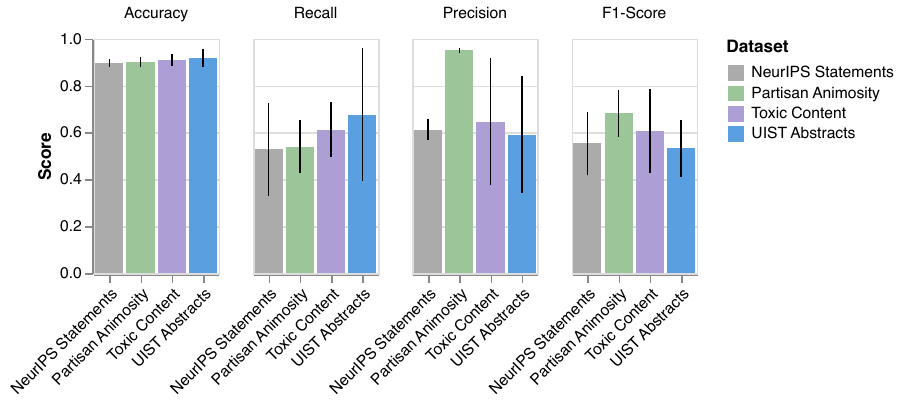}
  \caption{
    \textbf{Concept Classification Metrics}.
    Across the four LLooM scenario datasets, we observe high accuracy. There is substantial variance in classification performance both across datasets and across concepts within a dataset.
  }
  \label{fig:concept_score_eval}
  \Description{The figure displays four sets of bar charts: one for each metric of Accuracy, Recall, Precision, and F1 Score. Within each chart, the x-axis shows the dataset (NeurIPS statements, Partisan animosity, Toxic content, or UIST abstracts), and the y-axis shows the metric value. For accuracy scores, all dataset bars are high near 0.9, and confidence intervals are small. For recall, precision, and F1 score, the dataset bars are lower between 0.5 and 0.7 but differ greatly among different datasets, and the confidence intervals are very large for many bars.}
\end{figure*}

For classification metrics across datasets, we observe a mean accuracy of $0.91$, precision of $0.70$, recall of $0.59$, and F1 score of $0.59$; per-dataset metric results are shown in Figure~\ref{fig:concept_score_eval} and Table~\ref{tab:concept_score_eval}. Given that the concepts in this set are quite complex, and given that the documents are relatively long text examples, the scoring procedure achieves relatively strong performance results. However, this performance varies quite widely both across datasets and across concepts within a dataset.

To provide a point of comparison on this variability, we calculated inter-rater reliability between LLooM and each human annotator as well as between the two human annotators (A1 and A2). 
Across the four concepts, Cohen's $\kappa$ between the two human annotators was 0.64; meanwhile, the IRR between LLooM and A1 was 0.63, and the IRR between LLooM and A2 was 0.645. Thus, LLooM's annotations perform quite comparably to that of other human annotators. Per-concept IRR values are reported in Table~\ref{tab:concept_score_eval_irr}.

\begin{table}[!tb]
\centering
\caption{
Per-Dataset Classification Metrics. We report means and standard deviations for classification metrics on each LLooM scenario dataset. We observe considerable variance in classification performance across concepts and datasets.}
\begin{tabular}{lcccc}
\toprule
\textbf{Dataset}   & \textbf{Accuracy} & \textbf{Precision} & \textbf{F1 Score} \\ 
\midrule
NeurIPS Statements & 0.90 (0.02) & 0.61 (0.05) & 0.55 (0.14)    \\
Partisan Animosity & 0.90 (0.02) & 0.95 (0.01) & 0.68 (0.10)   \\
Toxic Content      & 0.91 (0.02) & 0.65 (0.27) & 0.61 (0.18)   \\
UIST Abstracts     & 0.92 (0.04) & 0.59 (0.25) & 0.53 (0.12)   \\ 
\bottomrule 
\end{tabular}
\label{tab:concept_score_eval}
\end{table}

\begin{figure}[!bt]
  \includegraphics[width=\linewidth]{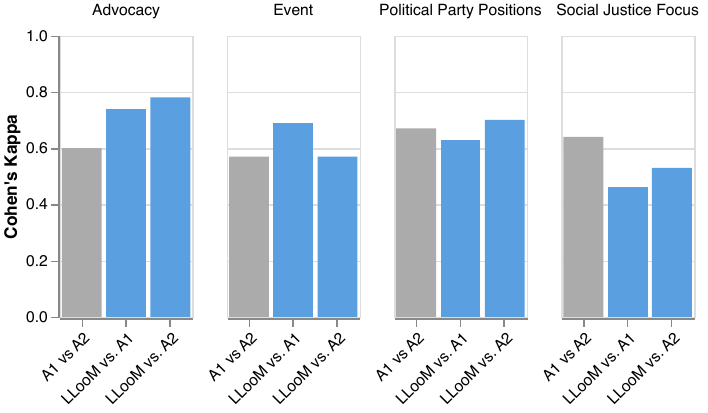}
  \caption{
    \textbf{Per-Concept Inter-rater Reliability}.
    Across the four concepts, we find similar, moderate-to-high Cohen's $\kappa$ values for the pair of human annotators (A1 and A2) and for LLooM when paired with each human annotator.
  }
  \label{fig:concept_score_eval_irr}
  \Description{The figure displays four sets of bar charts: one for each concept in the IRR Partisan Animosity dataset: Advocacy, Event, Political Party Positions, and Social Justice Focus. The x-axis of each chart has ``A1 vs. A2'', ``LLooM vs. A1'', and ``LLooM vs. A2'', and the y-axis displays Cohen’s Kappa. Across the charts, the Cohen’s Kappa values are very similar among the three bars.}
\end{figure}

\begin{table}[!tb]
\centering
\caption{
We observe that LLooM achieves inter-rater reliability levels (Cohen's $\kappa$) comparable to that of human annotators (A1 and A2). Agreement is moderate to high.
}
\begin{tabular}{p{4cm} p{1cm} p{1cm} p{1cm}}
\toprule
\textbf{Concept} &  \textbf{A1-A2} & \textbf{LLooM-A1} & \textbf{LLooM-A2} \\[0.1cm]
\midrule
{Advocacy}: \textit{Does the text example advocate for a cause or issue?} &    0.60 & 0.74 & 0.78 \\[0.5cm]
{Event}: \textit{Is this text example related to an event?} &     0.57 & 0.69 & 0.57 \\[0.2cm]
{Political Party Positions}: \textit{Does the text example mention the positions or actions of political parties?} &     0.67 & 0.63 & 0.70 \\[0.5cm]
{Social Justice Focus}: \textit{Does the text example emphasize working towards a just future?} &     0.64 & 0.46 & 0.53 \\[0.5cm]
\bottomrule
\end{tabular}
\label{tab:concept_score_eval_irr}
\end{table}

Qualitatively analyzing error cases where LLooM disagreed with human annotators, we find that the LLooM annotations generally appeared reasonable; they tended to be plausible, but differing, interpretations of the text. 
For false positives where LLooM marked documents as matching a concept while the human annotator (A1) did not, differences seemed to stem from differing \textit{thresholds} of concept matching. In general, LLooM was more likely to label examples as positive for a concept, especially for borderline cases. However, its decisions seem to fall within a a grey area of reasonability given the subjective nature of many of these concepts. For example, the following example was labeled by LLooM as positive for the \textit{Advocacy} concept while the human annotator marked the example as negative: ``Today was made possible because of the Pennsylvania Democrats who organized, knocked doors, donated, and voted.'' In this case, the text implicitly references causes or issues that are supported, but does not explicitly advocate for a cause. This subjectivity could reasonably lead to differing labels.

Meanwhile, for false negatives where the human annotator marked documents as matching a concept while LLooM did not, a common trend was that the examples required a deeper level of expertise or appreciation of nuance. This may be a failure mode for LLMs like GPT-3.5, which underlies the LLooM \texttt{Score} operator. For example, with the same \textit{Advocacy} concept above, the following example (excerpted) was labeled by the human annotator as positive while LLooM labeled the example as negative: ``[...] I will be working to make sure Head Start \& Early Head Start has the resources it needs to serve thousands of children in Middle GA.'' The text did not explicitly advocate for a cause or ask others to join with the typical language of advocacy, but it mentioned a particular government program (Head Start) that promotes school readiness for pre-school-age children from low-income families. The annotator had this knowledge and interpreted the text as advocating for this cause, while the LLM may not have had this context.

Overall, this evaluation and error analysis supports earlier evidence that LLooM performs annotation at a level comparable to that of another human annotator, but that it cannot avoid the inherent disagreement that will arise from subjective annotation tasks~\cite{disagreement_deconv}.

\begin{figure*}[!tb]
  \includegraphics[width=1.0\textwidth]{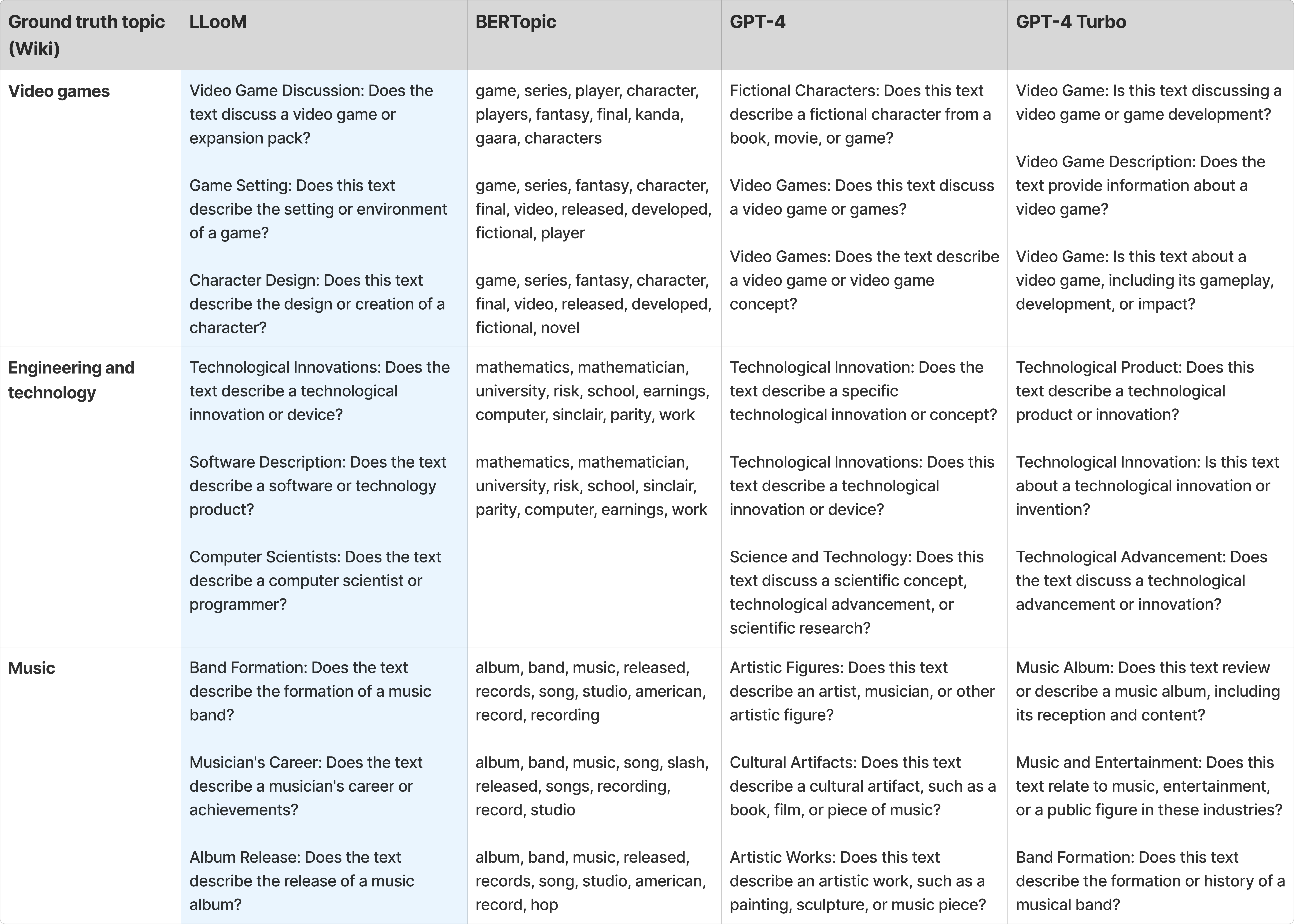}
  \caption{
    Sample of Wiki Dataset results for LLooM and baseline methods.
  }
  \label{fig:tech_eval_outputs_wiki}
  \Description{This figure displays a four-column table with LLooM, BERTopic, GPT-4, and GPT-4 Turbo. It has rows for different ground truth Wiki dataset topics: Video games, Engineering and technology, and Music. The cells are populated with examples of generated topics from each method that match to each ground truth topic.}
\end{figure*}

\begin{figure*}[!tb]
  \includegraphics[width=1.0\textwidth]{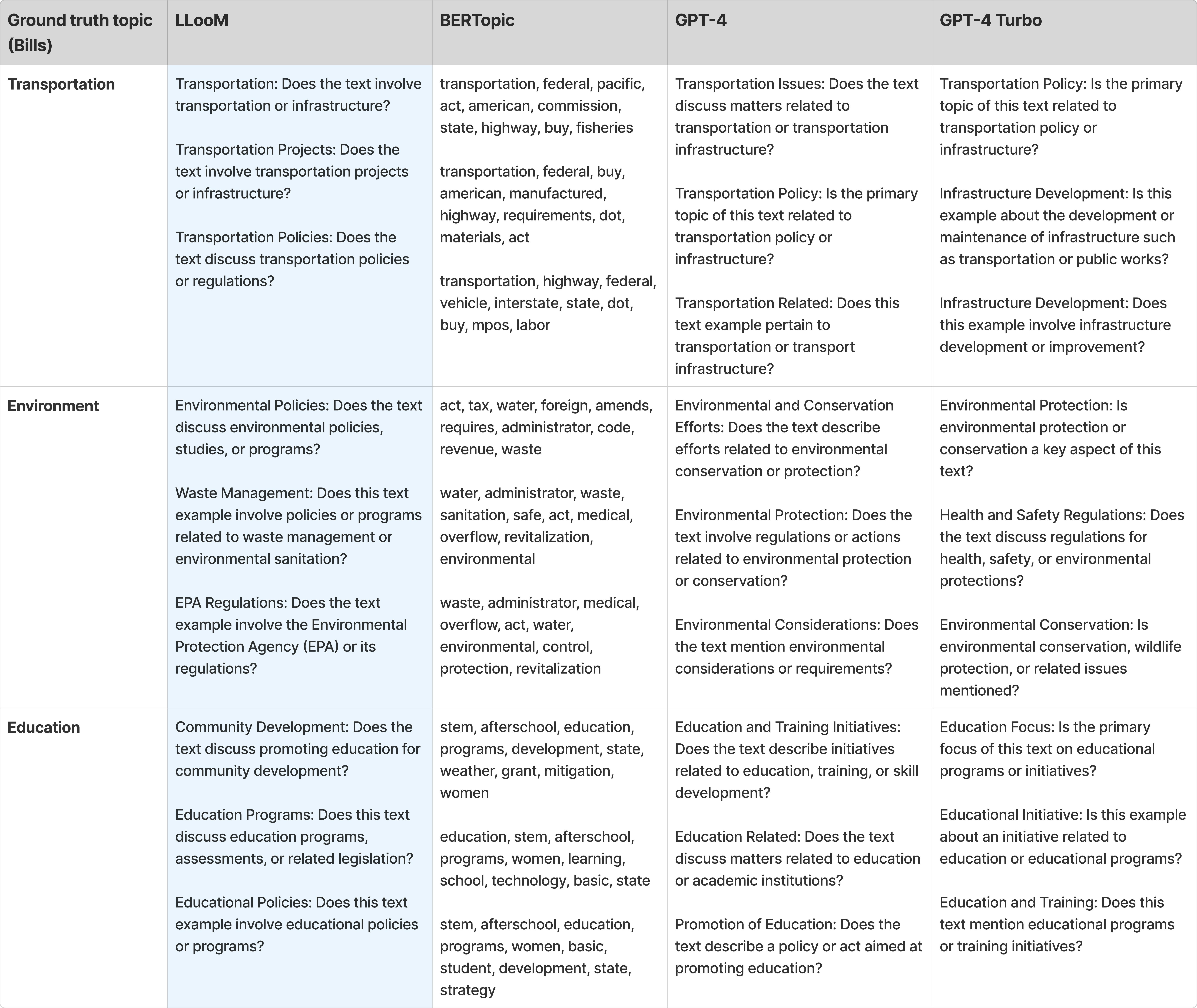}
  \caption{
    Sample of Bills Dataset results for LLooM and baseline methods.
  }
  \label{fig:tech_eval_outputs_bills}
  \Description{(Same table format as Figure~\ref{fig:tech_eval_outputs_wiki}) Here, the rows are different ground truth Bills dataset topics: Transportation, Environment, and Education.}
\end{figure*}

\begin{figure*}[!tb]
  \includegraphics[width=1.0\textwidth]{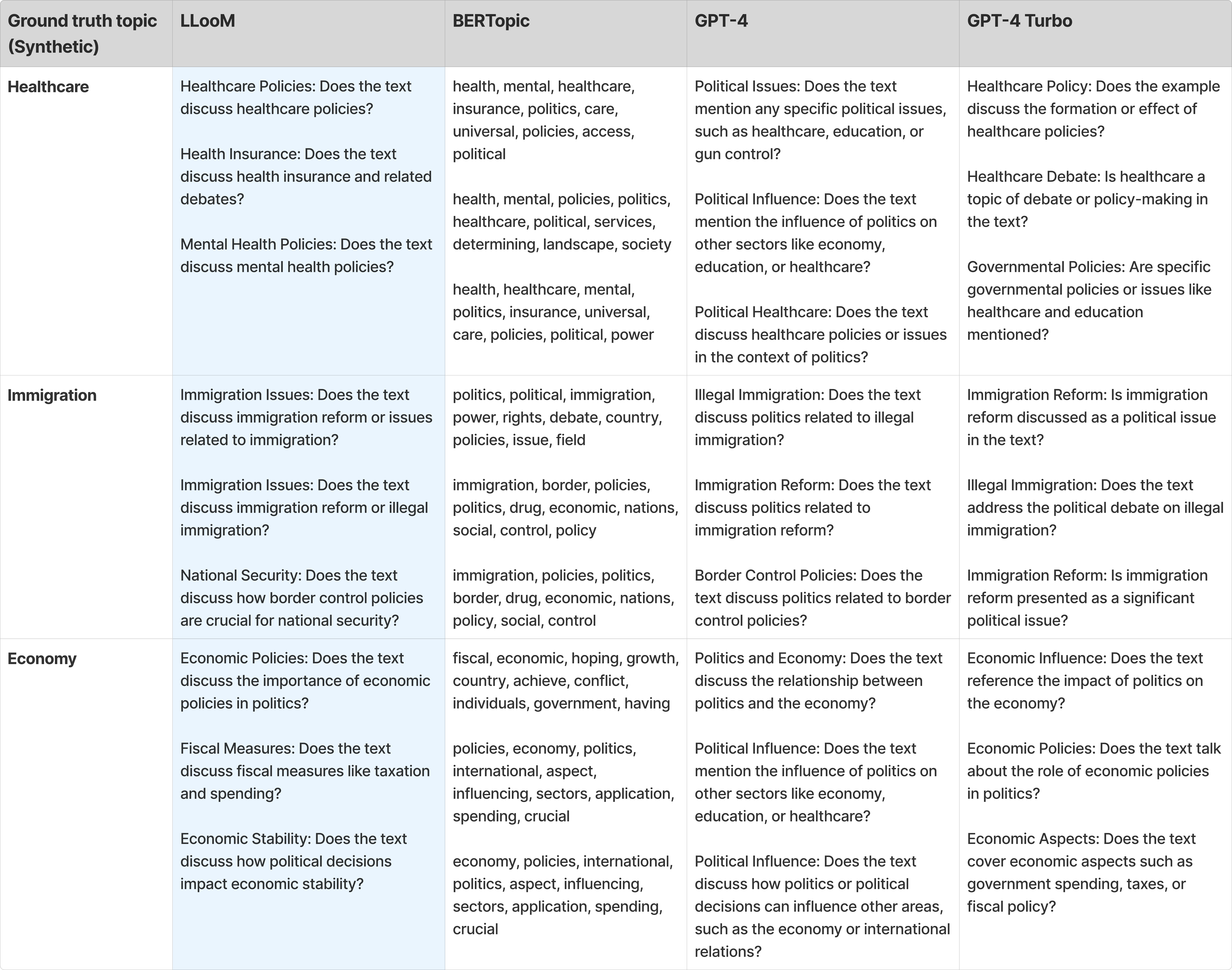}
  \caption{
    Sample of Synthetic Dataset results for LLooM and baseline methods.
  }
  \label{fig:tech_eval_outputs_synth}
  \Description{(Same table format as Figure~\ref{fig:tech_eval_outputs_wiki}) Here, the rows are different ground truth Synthetic dataset concepts: Healthcare, Immigration, and Economy.}
\end{figure*}

\subsection{Technical Evaluation: Concept Generation Outputs}
\label{appendix:tech_eval_outputs}
We include sample outputs for LLooM, BERTopic, GPT-4, and GPT-4-Turbo on the benchmark datasets (Wiki and Bills) and the synthetic dataset from the technical evaluation in Section~\ref{section:tech_eval}. For each dataset, we sampled three ground truth topics. Then, for each of the four methods, we sampled up to three generated concepts that matched the ground truth topic from across all trials. 
We display the results for the Wiki dataset in Figure~\ref{fig:tech_eval_outputs_wiki} for the ``Video games,'' ``Engineering and technology,'' and ``Music'' concepts.
We display the results for the Bills dataset in Figure~\ref{fig:tech_eval_outputs_bills} for the ``Transportation,'' ``Environment,'' and ``Education'' concepts.
We display the results for the synthetic dataset in Figure~\ref{fig:tech_eval_outputs_synth} for the ``Healthcare,'' ``Immigration,'' and ``Economy'' concepts.

\subsection{Technical Evaluation: Synthetic Dataset Concepts}
\label{appendix:synth_data_concepts}
To generate the synthetic data, we used the following set of 10 Generic concepts and 40 Specific concepts:
\begin{enumerate}
    \small
    \item \textbf{Generic}: Election Campaigns, \textbf{Specific}: Fundraising, Candidate Profiles, Political Rallies, Campaign Promises
    \item \textbf{Generic}: Government Policies, \textbf{Specific}: Healthcare Policies, Education Policies, International Relations Policies, Economic Policies
    \item \textbf{Generic}: Political Parties, \textbf{Specific}: Party Platforms, Party Leadership, Party History, Party Factionalism
    \item \textbf{Generic}: Human Rights, \textbf{Specific}: LGBTQ+ Rights, Women's Rights, Racial Equality, Children's Rights
    \item \textbf{Generic}: Immigration, \textbf{Specific}: Border Control Policies, Refugee Policies, Immigration Reform, Illegal Immigration
    \item \textbf{Generic}: Economy, \textbf{Specific}: Taxes, Unemployment, Fiscal Policy, Government Spending
    \item \textbf{Generic}: Healthcare, \textbf{Specific}: Universal Healthcare, Mental Health, Drug Policy, Health Insurance
    \item \textbf{Generic}: Environment, \textbf{Specific}: Climate Change, Renewable Energy, Nature Conservation, Air Pollution
    \item \textbf{Generic}: Foreign Policy, \textbf{Specific}: Trade Agreements, War and Peace, Diplomatic Relations, International Aid
    \item \textbf{Generic}: Gun Control, \textbf{Specific}: Background Checks, Assault Weapons Ban, Gun Control Legislation, Second Amendment Rights
\end{enumerate}

%TC:endignore

\end{document}